\documentclass[12pt]{article}
\usepackage{epsfig,amsmath,amssymb,mathrsfs}
\usepackage[utf8]{inputenc}
\usepackage[colorlinks=true,citecolor=blue,,linktocpage=true,linkcolor=blue,urlcolor=black]{hyperref}
\usepackage{mathtools,slashed}
\usepackage{array}
\usepackage{xcolor}
\usepackage[force]{feynmp-auto}
\usepackage{caption}
\usepackage{subcaption}

\usepackage[left=2.5cm,bottom=3cm,right=2.5cm,top=3cm]{geometry}

\usepackage{overpic,youngtab}
\usepackage[latin,english]{babel}
\usepackage{amsmath}
\usepackage{amssymb}
\usepackage{epstopdf}
\usepackage{graphics,psfrag,rotating}
\usepackage{mathabx}
\usepackage{graphicx}
\usepackage{dcolumn}
\usepackage{float}
\usepackage{pdflscape}
\usepackage{array}
\usepackage{booktabs}
\usepackage{amscd}
\usepackage{mathtools}
\usepackage{fancybox}
\usepackage{tensor}
\usepackage{fix-cm}
\usepackage[colorlinks=true,citecolor=blue,,linktocpage=true,linkcolor=blue,urlcolor=black]{hyperref}
\usepackage{cleveref}
\usepackage{tikz}
\usetikzlibrary{calc}
\usetikzlibrary{intersections}
\usetikzlibrary{positioning}
\usetikzlibrary{arrows}
\usetikzlibrary{shapes.misc}

\usetikzlibrary{patterns,snakes}
\tikzstyle{spring}=[line width=0.8,black,snake=coil,segment amplitude=4.25,segment length=4.75,line cap=round]
\usetikzlibrary{decorations.pathmorphing}
\usetikzlibrary{decorations.markings}
\usetikzlibrary{arrows.meta}
\usetikzlibrary{matrix}
\usetikzlibrary{positioning}
\usetikzlibrary{arrows}
\usepackage{titlesec}
\usepackage{abstract}
\usepackage{cancel}
\usepackage{mathtools}
\usepackage{pdflscape}
\usepackage{graphicx}
\definecolor{mypink}{HTML}{FDA4BA}
\usepackage{pdflscape}
\usepackage{subcaption}

  {\end{list}}%

\makeatletter
\renewcommand{\section}{\setcounter{equation}{0}\@startsection
 {section}%
 {1}%
 {0pt}%
 {-1\baselineskip}%
 {0.4\baselineskip}%
 {\bfseries\large}}%
\renewcommand{\subsection}{\@startsection
 {subsection}%
 {2}%
 {0pt}%
 {-0.75\baselineskip}%
 {0.2\baselineskip}%
 {\bfseries}}%
\renewcommand{\subsubsection}{\@startsection
 {subsubsection}%
 {3}%
 {0pt}%
 {-0.5\baselineskip}%
 {0.1\baselineskip}%
 {\sc}}%
\makeatother

\DeclareMathAlphabet{\mathpzc}{OT1}{pzc}{m}{it}

\def\be{\begin{equation}}
\def\ee{\end{equation}}
\def\a{\alpha}

\def\d{\delta}

\def\g5{\gamma_{5}}

\def\m{\mu}
\def\n{\nu}
\def\r{\rho}

\def\id3k{\int\!\! \dfrac{d^3\!\vec{k}}{(2\pi)^3 2E(\vec{k})}}

\def\idpD{\int\!\! \dfrac{d^{D}p\!}{(2\pi)^{D}}}

\newcommand{\bea}{\begin{eqnarray}}
\newcommand{\eea}{\end{eqnarray}}
\newcommand{\beann}{\begin{eqnarray*}}
\newcommand{\eeann}{\end{eqnarray*}}
\newcommand{\ba}{\begin{array}}
\newcommand{\ea}{\end{array}}

 \def\g {\gamma}

\newcommand{\email}[1]{\href{mailto:#1}{\tt #1}}

\begin{document}

\rightline{\scriptsize{FT/UCM 214-2023}}
\rightline{\scriptsize{IFT-UAM/CSIC-23-81}}
\vglue 50pt

\begin{center}

{\LARGE \bf The one-loop unimodular graviton propagator in any dimension}\\
\vskip 1.0true cm
{\Large Jesus Anero$^{\dagger}$, Carmelo P. Martin$^{\dagger\dagger}$ and Eduardo Velasco-Aja{$^{\dagger\dagger\dagger}$}}
\\
\vskip .7cm
{
	{$\dagger$,{$\dagger\dagger\dagger$}Universidad Auto\'noma de Madrid (UAM), Departamento de F\'isica Te\'orica and Instituto de F\'{\i}sica Te\'orica (IFT-UAM/CSIC),
    C/ Nicol\'as Cabrera 13-15 Universidad Auto\'noma de Madrid Cantoblanco, Madrid 28049, SPAIN}\\
	\vskip .1cm
	{$\dagger\dagger$Universidad Complutense de Madrid (UCM), Departamento de Física Teórica and IPARCOS, Facultad de Ciencias Físicas, 28040 Madrid, Spain}
	
	\vskip .5cm
	\begin{minipage}[l]{.9\textwidth}
		\begin{center}
			\textit{E-mail:}
			\email{$\dagger$jesusanero@gmail.es},
			\email{$\dagger\dagger$carmelop@fis.ucm.es}.
			{$\dagger\dagger\dagger$}\email{eduardo.velasco@uam.es}.
			
		\end{center}
	\end{minipage}
}
\end{center}
\thispagestyle{empty}

\begin{abstract}
For unimodular gravity, we work out{,} by using dimensional regularization{,} the complete  one-loop correction to the graviton propagator in any space-time dimension. The computation is carried out within the framework where unimodular gravity has Weyl invariance in addition to the transverse diffeomorphism gauge symmetry. Thus, no {Lagrange }multiplier is introduced to enforce the unimodularity condition.  The quantization of the theory is carried out by using the BRST framework and there considering a large continuous family of gauge-fixing terms. The BRST formalism is developed in such a way that the set of ghost, {anti-ghost} and auxiliary  fields and their BRST changes do not depend on the space-time dimension, as befits dimensional regularization.
As an application of our general result, and at $D=4$, we  obtain  the renormalized one-loop graviton propagator in the dimensional regularization minimal {subtraction} scheme. We do so by considering two simplifying gauge-fixing choices.
\end{abstract}

{\em Keywords:} Models of quantum gravity, unimodular gravity, renormalization.
\vfill
\clearpage

\section{Introduction.}
If one applies the effective field theory formalism to General Relativity, one concludes that the value of the Cosmological Constant should  be some hundred of orders of magnitude larger than its measured value. This a part of the so-called Cosmological Constant problem \cite{Weinberg:1988cp}. Unimodular gravity furnishes a solution \cite{vanderBij:1981ym, Zee:1983jg, Buchmuller:1988wx, Henneaux:1989zc, Buchmuller:2022msj} to this part of the problem since in that gravity theory the vacuum energy does not gravitate: in unimodular gravity the metric {always has}  determinant equal to $-1$ --see \cite{Carballo-Rubio:2022ofy, Alvarez:2023utn}, for introductions.

There are several approaches to define unimodular gravity as a quantum field theory, {e.g.}, \cite{Eichhorn:2013xr, Padilla:2014yea, Alvarez:2015sba, Bufalo:2015wda,  deLeonArdon:2017qzg, DeBrito:2019gdd, Baulieu:2020obv, deBrito:2021pmw, Kugo:2022iob}. It is not known whether they yield the same quantum theory, even when the background metric is Minkowski, for they involve different sets of {ghosts}. This is an open problem, as it is their equivalence to General Relativity when the Cosmological Constant is set to zero.

In this paper we shall adhere to the formulation of unimodular gravity introduced in \cite{Alvarez:2006uu, Alvarez:2005iy}, which was put into BRST form in  \cite{Alvarez:2015sba}. The starting point of this formulation is to solve first the unimodularity constraint ${\rm det}\hat{g}_{\mu\nu}(x)=-1$ --$\hat{g}_{\mu\nu}$ being the unimodular metric in $D$-dimensional space-time- by  introducing an unconstrained tensor field $g_{\mu\nu}(x)$ such that
\begin{equation*}
\hat{g}_{\mu\nu}(x)=\dfrac{g_{\mu\nu}(x)}{|g|^{1/D}(x)}.
\end{equation*}
Then, the path integral of the theory is defined by using the standard linear splitting $g_{\mu\nu}(x)=\bar{g}_{\mu\nu}+h_{\mu\nu}$, along with BRST techniques. $\bar{g}_{\mu\nu}$ is the background field and $h_{\mu\nu}$ is the graviton field, which is integrated over in the path integral. Of course, one may use any other type of splitting --{e.g.}, the splitting induced by the exponential parametrization-- to define the quantum theory. Let us stress that in general the linear splitting cannot be used in a given formulation of unimodular gravity which {} the unimodularity constraint by using a Lagrange multiplier \cite{Padilla:2014yea, Bufalo:2015wda, Baulieu:2020obv, Kugo:2022iob}, say $\lambda(x)$. Indeed, as shown in \cite{Alvarez:2023kuw},  UV divergent contributions which are quadratic in $\lambda(x)$, arise at  one and higher loop levels. As also shown in \cite{Alvarez:2023kuw} such a problem with the use of a Lagrange multiplier never happens if the  parametrization of the metric in terms of the  graviton field, $h_{\mu\nu}$,  is such that the unimodularity condition is equivalent to $h_{\mu\nu}$ being traceless --an instance of such parametrization is the exponential parametrization. Because we prefer to have any kind of sensible splitting at our disposal, we favour in this paper the formulation introduced in \cite{Alvarez:2006uu, Alvarez:2005iy}. Another reason may be found in that this formulation, which has Weyl invariance, finds nice accommodation within the string theory framework \cite{Garay:2023nco}. Further, the formulation in question also fits in the gauge/gravity duality framework \cite{Anero:2022rqi}.

One of the most important quantities in a quantum field theory of {gravity} is the complete one-loop graviton propagator. Indeed, this quantity gives rise, in any space-time dimension, to a highly {non-trivial} one-loop contribution to the scattering of two particles; which in turn yields a part of the quantum correction to Newton's Law \cite{Donoghue:1994dn}. It is also advisable to compute that one-loop contribution for large families of gauge-fixing choices, since the cancellation of the gauge dependence when computing a physical quantity is a non-trivial check that  our computation has been done correctly. In the General Relativity case, such  computations were done long ago in \cite{Capper:1978yf, Capper:1979ej} and just last year \cite{Brandt:2022und}. These {types} of computations --perhaps because they are far more involved than in General Relativity-- have never been done for unimodular gravity. The final goal of this paper is to remedy this state of affairs and compute the complete one-loop correction to the graviton propagator for a very large continuous family of gauge-fixing terms in any number of space-time dimensions. To do so we shall adapt the BRST formalism in \cite{Alvarez:2015sba} to the case of a Minkowski {background }and generalize the choice of gauge-fixing term. Then,  we shall work out the free propagators and the interaction vertices involving three fields. This, in the case at hand, is a very laborious task, let alone the  computation of the one-loop contributions, that we have been able to do thanks to invaluable help from the symbolic manipulation systems  ${\it FORM}$  \cite{Ruijl:2017dtg}, ${\it xAct}$\cite{xAct} and ${\it Mathematica}$\cite{Mathematica}. Let us point out that our BRST ghost, anti-ghost and auxiliary field system and their transformations do not depend on the space-time dimension so that we can keep $D$ arbitrary in our computations. This independence on the space-time dimension does not hold for the BRST quantization approach of reference \cite{Kugo:2022iob}.

The {layout} of the paper is as follows. In section 2, we carry out the BRST quantization of our theory, introducing a continuous large family of gauge-fixing terms. The free propagators, relevant vertices and Feynman rules are displayed in section 3. Section 4 is devoted to the computation, in dimensional regularization, of the one-loop contribution to the graviton propagator in any number of dimensions. In this section two choices of gauge parameters that greatly simplify the final result are also introduced. Comments and conclusions can be found in section 5. We include three appendices. In {Appendix} A, we give formulae leading to the renormalized, at $D=4$, one-loop graviton propagator in the minimal subtraction scheme of dimensional regularization; here some choices of {gauge parameters} are put in place. The value of each non-vanishing 1PI Feynman diagram contributing to the one-loop graviton propagator is given in Appendix B.
In {Appendix} C, we show how our general free gravity propagator yields Newton's law and discuss how this fact demands that two gauge-dependent coefficients of the free propagator satisfy a precise linear equation.

\section{BRST quantization of unimodular gravity with Weyl invariance.}

The classical action of our formulation of unimodular gravity  reads
\begin{equation}
 S_{\text{\tiny{UG}}}\,=\, -2\int d^D x\,R[\hat{g}_{\mu\nu}],
 \label{UGaction}
\end{equation}
where $R[\hat{g}]$ is the Ricci scalar for the unimodular metric $\hat{g}_{\mu\nu}$. The metric $\hat{g}_{\mu\nu}$ is defined in terms of the unconstrained tensor field $g_{\mu\nu}$ as follows:
\begin{equation}
\hat{g}_{\mu\nu}(x)=\frac{g_{\mu\nu}(x)}{|g|^{1/D}(x)}.
\label{gravitonfield}
\end{equation}
$g_{\mu\nu}$ is a tensor field under transverse diffeomorphisms and it is the field variable that will be used, being unconstrained, to quantize the classical theory. $D$ is the space-time dimension.

The action in (\ref{UGaction}) as a functional of $g_{\mu\nu}$ has got two gauge symmetries. These gauge symmetries are transverse diffeomorphisms and Weyl transformations; which, in infinitesimal form, read
\begin{equation}
\begin{array}{l}
{\delta_{\rm \tiny{TD}}g_{\mu\nu}(x)=\nabla_\mu\epsilon^{T}_\nu(x)+\nabla_\nu\epsilon^{T}_\mu(x),\quad \epsilon^{T}_\mu(x)=g_{\mu\nu}\epsilon^{T\,\nu}(x),\quad\partial_{\mu}\epsilon^{T\,\mu}(x)=0,}\\[4pt]
{\delta_{W} g_{\mu\nu}(x)=2 \epsilon(x) g_{\mu\nu}(x).}
\end{array}
\label{gaugesym}
\end{equation}
In the previous definitions $\epsilon^{T\,\mu}(x)$  and $\epsilon(x)$ are  infinitesimal vector and scalar fields, respectively, and $\nabla_\mu$ is the covariant derivative for  $g_{\mu\nu}$.

With the goal of formulating a field theory of gravitons propagating on Minkowski space-time, we shall consider the standard linear splitting of $g_{\mu\nu}$ in (\ref{gravitonfield}):
\begin{equation*}
g_{\mu\nu}(x)=\eta_{\mu\nu}+h_{\mu\nu}(x).
\end{equation*}
The field $h_{\mu\nu}$ is the graviton field.

To quantize the classical theory with action in (\ref{UGaction}) and gauge symmetries in (\ref{gaugesym}) we shall use the Batalin-Vilkovisky BRST formalism \cite{Batalin:1983wj}, for the vector field $\epsilon^{T\,\mu}$ above is constrained by transversality. This constraint makes the transverse diffeomorphism transformations in (\ref{gaugesym}) a first-stage gauge reducible
transformations, so that, according to \cite{Batalin:1983wj}, the following set of fields is to be introduced to set up the BRST formalism for the transverse diffeomorphism gauge symmetry:
\begin{equation}
\begin{array}{l}
{h_{\m\n}^{(0,0)},\,c^{(1,1)\,\mu},\,b^{(1,-1)\,\mu},\,f^{(0,0)\,\mu},\,\phi^{(0,2)},}\\[4pt]
{\pi^{(1,-1)},\,\pi'^{(1,1)},\,\bar{c}^{(0,-2)}\quad\text{and}\quad c'^{(0,0)}.}
\label{brsfieldsdiff}
\end{array}
\end{equation}

 On the other hand, the Weyl transformations in (\ref{gaugesym}) give rise to an irreducible gauge symmetry which demands the use of the following ghost, anti-ghost and auxiliary fields
\begin{equation}
c^{(1,1)},\,b^{(1,-1)}\quad\text{and}\text\quad f^{(0,0)},
\label{brsfieldsweyl}
\end{equation}
to formulate the BRST formalism. In the previous equations the first entry, $n$, of the superscript pair $(n,m)$ in the fields  is the Grassmann number; the second entry, $m$, is the ghost number.

Let us introduce the BRST operators, $s_{\rm \tiny{TD}}$, and $s_{\rm \tiny{W}}$, defining the BRST transformations associated to the transverse diffeomorphisms and Weyl transformations in (\ref{gaugesym}), respectively. Following ref.  \cite{Alvarez:2015sba}, we define the action of $s_{\rm \tiny{TD}}$ and $s_{\rm \tiny{W}}$ on the fields in (\ref{brsfieldsdiff}) and (\ref{brsfieldsweyl}) as follows
\begin{equation}
\begin{array}{l}
{s_{\rm \tiny{TD}} \eta_{\mu\nu}=0,}\\[4pt]
{s_{\rm \tiny{TD}} h_{\mu\nu}=\partial_\mu c^T_\nu+\partial_\nu c^T_\mu+c^{\rho T}\partial_\rho h_{\mu\nu}+h_{\rho\nu}\partial_\mu c^{T\rho}+h_{\rho\mu}\partial_\nu c^{T\rho},}\\[4pt]
{s_{\rm \tiny{TD}}c^{(1,1)\,\mu}=\frac{1}{\Box}\left(c^{\rho T}\partial_\rho c^{T\mu}\right)+\partial^\mu \phi^{(0,2)},}\\[4pt]
{s_{\rm \tiny{TD}}b^{(1,-1)\,\mu}=f^{(0,0)\,\mu},\quad
s_{\rm \tiny{TD}}f^{(0,0)\,\mu}=0,\quad
s_{\rm \tiny{TD}}\phi^{(0,2)}=0,}\\[4pt]
{s_{\rm \tiny{TD}}\pi^{(1,-1)}=0,\quad
s_{\rm \tiny{TD}}\pi'^{(1,1)}=0,\quad
s_{\rm \tiny{TD}}\bar{c}^{(0,-2)}=\pi^{(1,-1)},\quad
s_{\rm \tiny{TD}}c'^{(0,0)}=\pi'^{(1,1)},}\\[4pt]
{s_{\rm \tiny{TD}}c^{(1,1)}=c^{T\rho}\partial_\rho c^{(1,1)},\quad
s_{\rm \tiny{TD}}b^{(1,-1)}=c^{T\rho}\partial_\rho b^{(1,-1)},\,
s_{\rm \tiny{TD}}f^{(0,0)}=c^{T\rho}\partial_\rho f^{(0,0)},}\\[8pt]
{s_{\rm \tiny{W}} \eta_{\mu\nu}=0,}\\[4pt]
{s_{\rm \tiny{W}} h_{\mu\nu}=2c^{(1,1)}\left(\eta_{\mu\nu}+h_{\mu\nu}\right),}\\[4pt]
{s_{\rm \tiny{W}}c^{(1,1)\,\mu}=0,\quad
s_{\rm \tiny{W}}b^{(1,-1)\,\mu}=0,\quad s_{\rm \tiny{W}}f^{(0,0)\,\mu}=0,\quad s_{\rm \tiny{W}}\phi^{(0,2)}=0,}\\[4pt]
{s_{\rm \tiny{W}}\pi^{(1,-1)}=0,\quad
s_{\rm \tiny{W}}\pi'^{(1,1)}=0,\quad
s_{\rm \tiny{W}}\bar{c}^{(0,-2)}=0,\quad
s_{\rm \tiny{W}}c'^{(0,0)}=0,}\\[4pt]
{s_{\rm \tiny{W}}c^{(1,1)}=0,\quad
s_{\rm \tiny{W}}b^{(1,-1)}=f^{(0,0)},\quad
s_{\rm \tiny{W}}f^{(0,0)}=0.}
\label{soperators}
\end{array}
\end{equation}
In the previous equations indices are raised and lowered by using $\eta_{\mu\nu}$ and the following definition is assumed
\begin{equation*}
c^{ T\,\mu}=\left(\delta^\mu_\nu\Box-\partial^\mu\partial_\nu\right)c^{(1,1)\,\nu}.
\end{equation*}
Note that both $s_{\rm \tiny{TD}}$ and $s_{\rm \tiny{W}}$ have got both Grassmann number and ghost number equal 1.

Using the results
\begin{equation*}
\partial_\m\left(c^{\r T}\partial_\r c^{T\m}\right)=0,\quad \partial_\m\left[\frac{1}{\Box}\left(c^{\r T}\partial_\r c^{T\mu}\right)\right]=0
\end{equation*}
and the definitions in (\ref{soperators}), one may show that
\begin{equation*}
s_{\rm \tiny{TD}}^2=0,\quad s_{\rm \tiny{W}}^2=0\quad\text{and that}\quad\{s_{\rm \tiny{TD}},s_{\rm \tiny{W}}\}=0.
\end{equation*}

Next, we define the full BRST operator, $s$, as follows
\begin{equation}
s= s_{\rm \tiny{TD}}\,+\,s_{\rm \tiny{W}}.
\label{fullBRSTop}
\end{equation}

We are now ready to introduce the BRST invariant action $S_{\rm BRST}$ of our unimodular gravity theory with all the gauge symmetries in (\ref{gaugesym}) fixed:
\begin{equation}
S_{\rm BRST}\,=\,S_{\rm UG}\,-\,\int\,d^D x\, sX.
\label{BRSTaction}
\end{equation}
Note that $S_{\rm UG}$ is given in (\ref{UGaction}) and the integration domain is Minkowski space-time. $X$ is a local functional of the fields in (\ref{brsfieldsdiff}) and (\ref{brsfieldsweyl}), and their derivatives, such that ${\it i})$ it has Grassmann number equal to 1 and ghost number equal to -1, and ${\it ii})$ the differential operator in the contribution to $S_{\rm BRST}$ which is quadratic in the fields is invertible --this way free-propagators will exist. The operator $s$ is defined in (\ref{fullBRSTop}).

Our choice of $X$ will  be quadratic in the fields but general  enough as to introduce a large continuous gauge parameter  dependence in  most coefficients  of free propagator of the graviton --see (\ref{propsymbols}) and (\ref{Gpropaone}). Actually our choice of $X$ guarantees that one of the coefficients of two of the most involved terms, namely,
\begin{equation*}
\frac{1}{(p^2)^2}\left(\eta_{\mu_1\mu_3}p_{\mu_2} p_{\mu_4}+\eta_{\mu_1\mu_4}p_{\mu_2} p_{\mu_3}+\eta_{\mu_2\mu_3}p_{\mu_1} p_{\mu_4}+\eta_{\mu_2\mu_4}p_{\mu_1} p_{\mu_3}\right)\quad\text{and}\quad\frac{1}{(p^2)^3} p_{\m_1} p_{\m_2} p_{\m_3} p_{\m_4},
\end{equation*}
can always be set to zero by a suitable choice of the gauge parameters. Agreement with Newton's law makes it impossible { for} the existence of a gauge-fixing term such that the coefficients of both terms vanish --see Appendix C.

 Our $X$ is the following large family of functions:
\begin{equation}
X\,=\,X_{\rm \tiny{TD}}+\,X_{\rm \tiny{W}},
\label{OURX}
\end{equation}
where
\begin{equation}
\begin{array}{l}
{X_{\rm \tiny{TD}}=b_\mu\left[\gamma_1\partial^\nu h_\nu^\mu+\gamma_2\partial^\mu h+\rho_1 f^{\mu}\right]+\bar{c}\left[\alpha_2\partial^\mu c_\m +\frac{1}{2}\Box\pi'\right]+c'\left[\alpha_1 \partial^\mu
b_\mu-\frac{1}{2}\Box\pi\right],}\\[4pt]
{ X_{\rm \tiny{W}}=\partial_\mu b\,\partial^\mu\left(f-\alpha h\right).}
\label{XTDXW}
\end{array}
\end{equation}
$h$ stands for the contraction $h^{\mu}_{\mu}$. The fields in (\ref{XTDXW}) are the fields in (\ref{brsfieldsdiff}) and (\ref{brsfieldsweyl}) without the superscript pair $(m,n)$ so as to make the expressions more readable.
$X_{\rm \tiny{TD}}$ carries out the gauge-fixing of the transverse diffeomorphisms. $X_{\rm \tiny{W}}$ is needed to gauge-fix the invariance under Weyl transformations. $\gamma_1, \gamma_2, \alpha_1, \alpha_2$ and $\alpha$ are the gauge-fixing parameters defining the family of gauge-fixing terms in $S_{\rm BRST}$. These gauge-fixing parameters can have any value which yields well-defined free propagators. We shall see below that this is achieved if $\gamma_1, \alpha_1, \alpha_2$ and $\alpha$ never vanish. Finally, notice that $X$ in
(\ref{OURX}) is reminiscent of the  gauge-fixing terms  used in references \cite{Capper:1978yf, Capper:1979ej, Brandt:2022und} for General Relativity and generalizes quite a lot the gauge-fixing term used in \cite{Alvarez:2015sba} in the unimodular gravity case.

We shall see that all the free propagators obtained from $S_{\rm BRST}$ as defined by (\ref{BRSTaction}) and (\ref{OURX}) are massless. Hence, to compute the one-loop correction to the graviton propagator we may drop any contribution to $S_{\rm BRST}$ involving {more than} three fields. Indeed, the vertices involving four fields give rise to tadpole contributions, which vanish in dimensional regularization, and one cannot draw a one-loop propagator Feynman diagram with  vertices having more than four legs. Accordingly, let us express $S_{\rm BRST}$ as follows
\begin{equation*}
S_{\rm BRST}\,=\,S_{2}^{(\rm even)}+S_{2}^{(\rm odd)}+S_3^{(\rm even)}+S_3^{(\rm odd)}+S_{4}.
\end{equation*}
$S_{2}^{(\rm even)}$ and $S_{2}^{(\rm odd)}$ are the quadratic contributions with fields having even and odd Grassmann number, respectively.
$S_3^{(\rm even)}$ is the three-field contribution with fields having only even Grassmann number, $S_3^{(\rm odd)}$ the contribution involving three fields some of them having odd Grassmann number and $S_{4}$ the summand with more than three fields. A little algebra, with a lot of help from ${\it xAct}$, yields
\begin{equation}
\begin{array}{l}
{S_{2}^{(\rm even)}=  -\frac{1}{2}\int
d^D x\Big\{h_{\mu\nu}\bar{\Box}h^{\mu\nu}-2h^{\mu\lambda}\partial_\mu \partial_\nu h^\nu_\lambda-\frac{D+2}{D^2}h\bar{\Box}h+\frac{4}{D}h^{\mu\nu}\partial_\mu\partial_\nu h}\\[4pt]
{\phantom{S_{2}}
-2h^{\mu\nu}\Big[\frac{\gamma_1}{2}\left(\partial_\nu f_\mu+\partial_\mu f_\nu\right) +\gamma_2\eta_{\mu\nu}\partial^\lambda f_\lambda\Big]
+2\rho_1 f_\mu f^\mu
+2\alpha_1c'\partial^\mu f_\mu-2f \Box f +2\alpha h\Box f\Big\},}\\[14pt]
{S_2^{(\rm odd)}=\int d^Dx\Big\{b_\mu\Big[\gamma_1\Box c^{T\,\mu}+2\left(\gamma_1+D\gamma_2\right)\partial^\mu c\Big]-b_\mu\alpha_1\partial^\mu \pi'
-\pi\alpha_2\partial^\mu c_\mu-\pi\Box \pi'-2D\alpha\partial_\mu b\partial^\mu c\Big\}, }\\[14pt]
{S_{3}^{(\rm even)}= \int
d^D x \Big\{\frac{D+2}{D^2}\partial_\mu hh^{\rho\sigma}\partial^\mu h_{\rho\sigma}-\frac{D+2}{2D^3}h\partial_\mu h\partial^\mu h+\frac{2}{D^2}h\partial_\mu h\partial_\rho h^{\mu\rho}-\frac{1}{2}h^{\mu\nu}\partial_\mu h_{\rho\sigma}\partial_\nu h^{\rho\sigma}}\\[4pt]
{\phantom{S_{3}^{(\rm even)}= \int d^D x \Big\{}
+\frac{D+2}{2D^2}h^{\mu\nu}\partial_\mu h\partial_\nu h-\frac{2}{D}\partial^\mu hh^\rho_\sigma\partial^\sigma h_{\rho\mu}
-\frac{2}{D}h^{\mu\nu}\partial_\mu h\partial_\rho h_\nu^\rho-\frac{1}{D}h\partial^\mu h_{\rho\sigma}\partial^\sigma h^\rho_\mu}\\[4pt]
{\phantom{S}
+\frac{1}{2D}h\partial_\mu h^{\rho\sigma}\partial^\mu h_{\rho\sigma}-\frac{2}{D}h^{\mu\nu}\partial_\rho h_{\mu\nu}\partial_\sigma h^{\rho\sigma}+2h^{\mu\nu}\partial_\nu h_{\rho\sigma}\partial^\sigma h^{\rho}_{\mu}
+h^{\mu\nu}\partial^\rho h_{\nu\sigma}\partial^\sigma h_{\mu\rho}
-h^{\mu\nu}\partial_\rho h_{\nu\sigma}\partial^\rho h_
\mu^\sigma\Big\}}\\[4pt]
{\text{and}}\\[4pt]
{S_3^{(\rm odd)}=
\int d^Dx\Big\{-\gamma_1\partial_\nu b_\mu\left(c^{\rho T}\partial_\rho h^{\mu\nu}+h^{\rho\nu}\partial^\mu c^T_\rho+h^{\rho\mu}\partial^\nu c^T_\rho\right)-\gamma_2\partial^\mu b_\mu\left(c^{\rho T}\partial_\rho h+2h_{\rho\lambda}\partial^\lambda c^{T\rho}\right)}\\[4pt]
{\phantom{S_3^{(\rm odd)}=\int d^nx\Big\{}
-2\gamma_1 c\, h^{\mu\nu}\partial_\nu b_\mu-2\gamma_2 c\, h\partial^\mu b_\mu
+2\alpha\Box b\, c\, h+\alpha \partial_\rho b\, c^{T\rho}\Box h+\alpha \Box b\, c^{T\rho}\partial_\rho h}\\[4pt]
{\phantom{S_3^{(\rm odd)}=\int d^nx\Big\{}
+2\alpha \Box b\, \partial^\lambda c^{T\rho} h_{\rho\lambda}-\partial_\rho b\, c^{T\rho}\Box f -\Box b\, c^{T\rho}\partial_\rho f\Big\}.}
\label{S2S3}
\end{array}
\end{equation}

Furnished with the action $S_{\tiny{\rm BRST}}$ above, one  constructs the quantum theory by using the standard path integral method.

\section{Propagators, Vertices and Feynman rules.}

In this section we shall display the free propagators, the three-field vertices and the corresponding Feynman rules of our unimodular gravity theory.

\subsection{Free propagators.}
The free propagators and Fourier transforms of the fields with even Grassman number are denoted by the following symbols:
\begin{equation}
\begin{array}{l}
{\langle h_{\mu_1\mu_2}(x)h_{\mu_3\mu_4}(y)\rangle_0=\idpD e^{-i p(x-y)} iG^{(hh)}_{\mu_1\mu_2\mu_3\mu_4}(p),}\\[4pt]
{\langle h_{\mu_1\mu_2}(x)f_{\mu_3}(y)\rangle_0=\idpD e^{-i p(x-y)}\,iG^{(hf_\mu)}_{\mu_1\mu_2\mu_3}(p),}\\[4pt]
{\langle h_{\mu_1\mu_2}(x)f(y)\rangle_0=\idpD e^{-i p(x-y)}iG^{(hf)}_{\mu_1\mu_2}(p),\;\langle h_{\mu_1\mu_2}(x)c'(y)\rangle_0=\idpD \d\,e^{-i p(x-y)}\,iG^{(hc')}_{\mu_1\mu_2}(p),}\\[4pt]
{\langle f_{\mu_1}(x)f_{\mu_2}(y)\rangle_0=\idpD e^{-i p(x-y)} iG^{(f_\mu f_\nu)}_{\mu_1\mu_2}(p),\;
\langle f_{\mu_1}(x)f(y)\rangle_0=\idpD e^{-i p(x-y)} iG^{(f_\mu f)}_{\mu_1}(p),}\\[4pt]
{\langle f_{\mu_1}(x)c'(y)\rangle_0=\idpD e^{-i p(x-y)} iG^{(f_\mu c')}_{\mu_1}(p),\;\langle f(x)f(y)\rangle_0=\idpD e^{-i p(x-y)}iG^{(ff)}(p),}\\[4pt]
{\langle f(x)c'(y)\rangle_0=\idpD e^{-i p(x-y)} iG^{(f c')}(p),\;\langle c'(x) c'(y)\rangle_0=\idpD e^{-i p(x-y)} iG^{(c' c')}(p).}
\label{propsymbols}
\end{array}
\end{equation}
The values of these propagators are obtained from $S^{(\rm even)}_2$ in (\ref{S2S3}) by using standard path integral methods. A very lengthy computation yields

\begin{equation}
\begin{array}{l}
{G^{(hh)}_{\mu_1\mu_2\mu_3\mu_4}(p)=}\\[2pt]
{\frac{1}{2}\frac{1}{p^2}\left(\eta_{\mu_1\mu_3}\eta_{\mu_2\mu_4}+\eta_{\mu_1\mu_4}\eta_{\mu_2\mu_3}\right)
-\frac{D^2\a^2-2D+4}{D^2(D-2)\a^2}\frac{1}{p^2}\eta_{\mu_1\mu_2}\eta_{\mu_3\mu_4}
+\frac{2}{D-2}\frac{1}{(p^2)^2}\left(\eta_{\mu_3\mu_4}p_{\mu_1} p_{\mu_2}+\eta_{\mu_1\mu_2}p_{\mu_3} p_{\mu_4}\right)}\\[4pt]
{\quad\quad\quad\quad\quad-\frac{\g_1^2-4\r_1}{2\g_1^2}\frac{1}{(p^2)^2}\left(\eta_{\mu_1\mu_3}p_{\mu_2} p_{\mu_4}+\eta_{\mu_1\mu_4}p_{\mu_2} p_{\mu_3}+\eta_{\mu_2\mu_3}p_{\mu_1} p_{\mu_4}+\eta_{\mu_2\mu_4}p_{\mu_1} p_{\mu_3}\right)}\\[4pt]
{\quad\quad\quad\quad\quad\quad\quad\quad\quad\quad\quad-\frac{4(\g_1^2+2(D-2)\r_1)}{(D-2)\g_1^2}\frac{1}{(p^2)^3} p_{\mu_1} p_{\mu_2} p_{\mu_3} p_{\mu_4},}\\[14pt]
{G^{(hf_\mu)}_{\mu_1\mu_2\mu_3}(p)=-\frac{i}{\g_1}\frac{1}{(p^2)^2}\Big[p^2\left(\eta_{\mu_1\mu_3}p_{\mu_2}+\eta_{\mu_2\mu_3}p_{\mu_1}\right)-2p_{\mu_1} p_{\mu_2} p_{\mu_3}\Big],}
\label{Gpropaone}
\end{array}
\end{equation}
\begin{equation}
\begin{array}{l}
{\phantom{\frac{1}{2}\frac{1}{p^2}\left(\eta_{\mu_1\mu_3}\eta_{\mu_2\mu_4}+\eta_{\mu_1\mu_4}\eta_{\mu_2\mu_3}\right)
-\frac{D^2\a^2-2D+4}{D^2(D-2)\a^2}\frac{1}{p^2}\eta_{\mu_1\mu_2}\eta_{\mu_3\mu_4}
+\frac{2}{D-2}\frac{1}{(p^2)^2}\left(\eta_{\mu_3\mu_4}p_{\mu_1} p_{\mu_2}+\eta_{\mu_1\mu_2}p_{\mu_3} p_{\mu_4}\right)}}\\[4pt]
{G^{(hf)}_{\mu_1\mu_2}(p)=\frac{1}{D\a}\frac{1}{p^2}\eta_{\m_1\mu_2},}\\[14pt]
{G^{(hc')}_{\mu_1\mu_2}(p)=\frac{(D^2\a^2+2D-4)\g_1+2(D-2)D\g_2}{(D-2)D^2\a^2\a_1}\frac{1}{p^2}\eta_{\mu_1\mu_2}-\frac{D}{D-2}\frac{\g_1}{\a_1}\frac{1}{(p^2)^2}p_{\mu_1} p_{\mu_2},}\\[14pt]
{G^{(f_\mu f_\nu)}_{\mu_1\mu_2}(p)=0,\quad G^{(f_\mu f)}_{\mu_1}(p)=0,\quad G^{(f_\mu c')}_{\mu_1}(p)=-\frac{i}{\a_1}\frac{1}{p^2}p_{\mu_1},}\\[14pt]
{G^{(f f)}(p)=0,\quad G^{(f c')}(p)=\frac{1}{D}\frac{\g_1+D\g_2}{\a\a_1}\frac{1}{p^2},}\\[14pt]
{G^{(c'c')}(p)=-\frac{(D^3\a^2-D^2\a^2-2D+4)\g_1^2-4(D-2)D\g_1\g_2-2(D-2)D^2(\g_2^2+\a^2\r_1)}{(D-2)D^2\a^2\a_1^2}\frac{1}{p^2}.}
\label{Gpropatwo}
\end{array}
\end{equation}

Next, the following symbols
\begin{equation}
\begin{array}{l}
{\langle  c_{\mu_1}(x)b_{\m_2}(y)\rangle_0=\idpD e^{-i p(x-y)} iG^{(c_\mu b_\nu)}_{\mu_1\mu_2}(p),\;
\langle c(x) b_{\mu_1}(y)\rangle_0=\idpD e^{-i p(x-y)}\,iG^{(c b_\mu)}_{\mu_1}(p),}\\[4pt]
{\langle  c^T_{\mu_1}(x)b_{\m_2}(y)\rangle_0=(\Box_x \delta_{\m_1}^\rho-\partial^x_{\m_1}\partial_x^\rho)\langle  c_{\rho}(x)b_{\m_2}(y)\rangle_0=\idpD e^{-i p(x-y)} iG^{(c^T_\mu b_\nu)}_{\mu_1\mu_2}(p),}\\[4pt]
{\langle \pi'(x) b_{\mu_1}(y)\rangle_0=\idpD e^{-i p(x-y)}\,iG^{(\pi' b_\mu)}_{\mu_1}(p),}\\[4pt]
{\langle c_{\mu_1}(x) b(y)\rangle_0=\idpD e^{-i p(x-y)} iG^{( c_\mu b)}_{\mu_1}(p),\;
\langle c(x) b(y)\rangle_0=\idpD e^{-i p(x-y)}\,iG^{(cb)}(p),}\\[4pt]
{\langle \pi'(x)b(y)\rangle_0=\idpD e^{-i p(x-y)}\,iG^{(\pi' b)}(p),}\\[4pt]
{\langle c_{\mu_1}(x)\pi(y)\rangle_0=\idpD e^{-i p(x-y)} iG^{( c_\mu\pi)}_{\mu_1}(p),\;
\langle c(x)\pi(y)\rangle_0=\idpD e^{-i p(x-y)}\,iG^{(c\pi)}(p),}\\[4pt]
{\langle \pi'(x)\pi(y)\rangle_0=\idpD e^{-i p(x-y)}\,iG^{(\pi' \pi)}(p),}
\label{propevensymbols}
\end{array}
\end{equation}
{stand for the free-propagators and  Fourier transform of the fields with an odd Grassmann number}. From $S_{2}^{(\rm odd)}$ in (\ref{S2S3}), one gets
\begin{equation}
\begin{array}{l}
{G^{(c_\mu b_\nu)}_{\m_1\m_2}(p)=\frac{1}{\g_1}\frac{1}{{(p^2)^3}}\left(p^2\eta_{\m_1\m_2}-p_{\m_1} p_{\m_2}\right)+\frac{1}{\a_1\a_2}\frac{1}{ p^2}\,p_{\m_1} p_{\m_2},\quad
G^{(c b_\mu )}_{\m_1}(p)=-i\frac{\g_1+D\g_2}{D\a\a_1\a_2}\frac{1}{ p^2}\, p_{\m_1}}\\[4pt]
{G^{(c^T_\mu b_\nu)}_{\m_1\m_2}(p)=-\dfrac{1}{\gamma_1 (p^2)^2}(p^2\eta_{\m_1\m_2}-p_{\m_1}p_{\m_2}),}\\[4pt]
{G^{(\pi' b_\mu)}_{\m_1}(p)=-\frac{i}{\a_2}\frac{1}{p^2}\,p_{\m_1},\quad
G^{(c_\mu b)}_{\m_1}(p)=0,\quad
G^{(c b)}(p)=-\frac{1}{2D\a }\frac{1}{p^2},\quad
G^{(\pi' b)}(p)=0}\\[4pt]
{G^{(c_\mu\pi)}_{\m_1}(p)=-\frac{i}{\a_1}\frac{1}{p^2}\,p_{\m_1},\quad
G^{(c\pi)}(p)=-\frac{\g_1+D\g_2}{D\a\a_1}\frac{1}{p^2},\quad
G^{(\pi' \pi)}(p)=0.}
\label{oddprop}
\end{array}
\end{equation}

\subsection{Three-field vertices.}

We now come to the three-field vertices. { By a Fourier transformation, one can recast $iS_3^{(\rm even)}$ in (\ref{S2S3}), into the form, }

\be  iS_3^{(\rm even)}=\int \prod_{i=1}^{3}\frac{d^D p_i}{(2\pi)^D}(2\pi)^D\d(p_1+p_2+p_3)\dfrac{1}{3!}\,V^{\m_1\n_1,\m_2\n_2,\m_3\n_3}{(p_1,p_2,p_3)}h_{\m_1\n_1}(p_1)h_{\m_2\n_2}(p_2)h_{\m_3\n_3}(p_3),\nonumber\ee
where the three-graviton vertex reads
\begin{equation}
\begin{array}{l}
{V^{\m_1\n_1,\m_2\n_2,\m_3\n_3}(p_1,p_2,p_3)=}\\[4pt]
{i\,3!\,\mathcal{S}\Big\{-\frac{D+2}{D^2}\Big[p_1p_3\eta^{\m_1\n_1}\eta^{\m_2\m_3}\eta^{\n_2\n_3}\Big]+
\frac{D+2}{2D^3}\Big[p_2\cdot p_3\eta^{\m_1\n_1}\eta^{\m_2\n_2}\eta^{\m_3\n_3}\Big]
-\frac{2}{D^2}\Big[p^{\m_3}_2p^{\n_3}_3\eta^{\m_1\n_1}\eta^{\m_2\n_2}\Big]}\\[4pt]
{+\frac{1}{2}\Big[p^{\m_1}_2p^{\n_1}_3\eta^{\m_2\m_3}\eta^{\n_2\n_3}\Big]
-\frac{D+2}{2D^2}\Big[p^{\m_1}_2p^{\n_1}_3\eta^{\m_2\n_2}\eta^{\m_3\n_3}\Big]+\frac{2}{D}\Big[p
^{\m_3}_1p^{\n_2}_3\eta^{\m_1\n_1}\eta^{\m_2\n_3}\Big]}\\[4pt]
{+\frac{2}{D}\Big[p^{\m_1}_2p^{\m_3}_3\eta^{\m_2\n_2}\eta^{\n_1\n_3}\Big]+\frac{1}{D}\Big[p^{\n_3}_2p^{\n_2}_3\eta^{\m_1\n_1}\eta^{\m_2\m_3}\Big]
-\frac{1}{2D}\Big[p_2p_3\eta^{\m_1\n_1}\eta^{\m_2\m_3}\eta^{\n_2\n_3}\Big]}\\[4pt]
{+\frac{2}{D}\Big[p^{\n_3}_2p^{\m_3}_3\eta^{\m_1\m_2}\eta^{\n_1\n_2}\Big]-2\Big[p^{\n_1}_2p^{\n_2}_3\eta^{\m_3\m_1}\eta^{\m_2\n_3}\Big]
-\Big[p^{\n_3}_2p^{\n_2}_3\eta^{\n_1\m_2}\eta^{\m_1\m_3}\Big]+\Big[p_2\cdot p_3\eta^{\n_1\m_2}\eta^{\n_2\n_3}\eta^{\m_3\m_1}\Big]\Big\}.}
\label{3gravvertex}
\end{array}
\end{equation}
Note that $\mathcal{S}$ is a shorthand for a double symmetrization, namely, a symmetrization with regard to each pair on indices $\m_1\n_1$, $\m_2\n_2$ and $\m_3\n_3$, first, and, then, a symmetrization with regard to all pairs $(p1,\m_1\n_1), (p_2, \m_2\n_2),  (p_3, \m_3\n_3)$. Symmetrization of $F[1,2,3,...,n]$ is carried out with the formula
\begin{equation*}
\sum_{\Sigma}\,\frac{1}{n!}\,F[\Sigma(1),\Sigma(2),...,\Sigma(n)],
\end{equation*}
where the sum runs over all permutations --generally denoted by $\Sigma$-- of $(1,2,...,n)$.

$S_3^{(\rm odd)}$ as defined in (\ref{S2S3}) gives rise to the {following} three-field interaction terms in momentum space

\begin{equation*}
\begin{array}{l}
{iS_3^{(\rm odd)}=
\int \prod_{i=1}^{3}\frac{d^D p_i}{(2\pi)^D}(2\pi)^D\d(-p_1+p_2+p_3)}\\[4pt]
{\Big[V_1^{\m_1,\m_2,\m_3\n_3}(p_1,p_2,p_3)\,b_{\m_1}(p_1)c_{\m_2}^T(p_2)h_{\m_3\n_3}(p_3)
+V_2^{\m_1,\m_3\n_3}(p_1,p_2,p_3)\,b_{\m_1}(p_1)c(p_2)h_{\m_3\n_3}(p_3)+}\\[4pt]
{\phantom{\Big[}
V_3^{\m_3\n_3}(p_1,p_2,p_3)\,b(p_1)c(p_2)h_{\m_3\n_3}(p_3)
+V_4^{\m_2,\m_3\n_3}(p_1,p_2,p_3)\,b(p_1)c^T_{\m_2}(p_2)h_{\m_3\n_3}(p_3)+}\\[4pt]
{\phantom{\Big[}
V_5^{\m_2}(p_1,p_2,p_3)\,b(p_1)c^T_{\m_2}(p_2)f(p_3)\Big],}
\end{array}
\end{equation*}
with
\begin{equation}
\begin{array}{l}
{V_1^{\m_1,\m_2,\m_3\n_3}(p_1,p_2,p_3)=-i\g_1\Big[\frac{1}{2}\left(p_1^{\n_3}p_3^{\m_2}\eta^{\m_1\m_3}+p_1^{\m_3}p_3^{\m_2}\eta^{\m_1\n_3}\right)+
\frac{1}{2}\left(p_1^{\n_3}p_2^{\m_1}\eta^{\m_2\m_3}+p_1^{\m_3}p_2^{\m_1}\eta^{\m_2\n_3}\right)}\\[4pt]
{\quad\quad\quad\quad+\frac{1}{2}p_1\cdot p_2\left(\eta^{\m_1\m_3}\eta^{\m_2\n_3}+\eta^{\m_1\n_3}\eta^{\m_2\m_3}\right)\Big]
-i\g_2\Big[p_1^{\m_1}p_3^{\m_2}\eta^{\m_3\n_3}+p_1^{\m_1}p_2^{\m_3}\eta^{\m_2\n_3}+p_1^{\m_1}p_2^{\n_3}\eta^{\m_2\m_3}\Big],}\\[14pt]
{V_2^{\m_1,\m_3\n_3}(p_1,p_2,p_3)=\g_1\left(p_1^{\n_3}\eta^{\m_1\m_3}+p_1^{\m_3}\eta^{\m_1\n_3}\right)+2\g_2p_1^{\m_1}\eta^{\m_3\n_3},}\\[14pt]
{V_3^{\m_3\n_3}(p_1,p_2,p_3)=-2i\a p_1^2\eta^{\m_3\n_3},}\\[14pt]
{V_4^{\m_2,\m_3\n_3}(p_1,p_2,p_3)=\a p_1^{\m_2}p_3^2\eta^{\m_3\n_3}-\a p_1^2p_3^{\m_2}\eta^{\m_3\n_3}-\a p_1^2\left(p_2^{\m_3}\eta^{\m_2\n_3}+p_2^{\n_3}\eta^{\m_2\m_3}\right),}\\[14pt]
{V_5^{\m_2}(p_1,p_2,p_3)=-p_1^{\m_2}p_3^2+p_3^{\m_2}p_1^2.}
\label{oddvertices}
\end{array}
\end{equation}
\newpage
\subsection{Feynman rules.}

The results in the previous subsections lead to the Feynman rules in. {\cref{fig:1,fig:2}} . The reader should {bear} in mind the definitions in (\ref{Gpropaone}), (\ref{Gpropatwo}),  (\ref{oddprop}), ({\ref{3gravvertex}) and (\ref{oddvertices}), when consulting those rules.
\begin{figure}[htbp]
    \centering
\begin{subfigure}[b]{0.8 \textwidth}
    \centering
    \begin{tikzpicture}[scale=3,
        decoration={coil,amplitude=4.25,segment length=4.75},anchor=base, baseline]
    \draw[line width=0.8,black,  postaction={decorate, decoration={markings, mark=at position 1 with {\arrow[fill=black, scale=1.5]{Triangle}} } }] (0.3,0)--(-0.54,0);
    \draw[spring] (-0.54,0) -- (-1.3,0);
    \filldraw [black] (-1,0.05) circle (0pt) node[anchor=south ]{$\tensor{h}{_{\mu_1}_{\nu_1}}(p) $};
    \filldraw [black] (0,0.05) circle (0pt) node[anchor=south ]{$f(-p) $};
    \filldraw [black] (-0.5,-0.1) circle (0pt) node[anchor=north ]{\large $p$};
    \filldraw [black] (1.7,0) circle (0pt) node[anchor=east ]{ $=i\,G^{(hf)}_{\mu_1\nu_1}(p) $};
    \end{tikzpicture}

\end{subfigure}
\begin{subfigure}[b]{0.8 \textwidth}
    \centering
    \begin{tikzpicture}[scale=3,
        decoration={coil,amplitude=4.25,segment length=4.75},anchor=base, baseline]
        \draw[line width=0.8,black,  postaction={decorate, decoration={markings, mark=at position 1 with {\arrow[fill=black, scale=1.5]{Triangle}} } }] (-0.54,0)--(-0.541,0);
        \draw[spring] (-.532,0) -- (-1.3,0);
        \draw[spring] (-.47,0) -- (.3,0);
        \filldraw [black] (-1,0.05) circle (0pt) node[anchor=south ]{$\tensor{h}{_{\mu_1}_{\nu_1}}(p) $};
        \filldraw [black] (0,0.05) circle (0pt) node[anchor=south ]{$\tensor{h}{_{\mu_2}_{\nu_2}}(-p) $};
        \filldraw [black] (-0.5,-0.1) circle (0pt) node[anchor=north ]{\large $p$};
       \filldraw [black] (2,0) circle (0pt) node[anchor=east ]{ $=i\, G^{(hh)}_{\mu_1\nu_1\mu_2\nu_2}(p)   $};
    \end{tikzpicture}

\end{subfigure}
\begin{subfigure}[b]{0.8 \textwidth}
    \centering
    \begin{tikzpicture}[scale=3,
        decoration={coil,amplitude=4.25,segment length=4.75},anchor=base, baseline]
    \draw[line width=0.8,black,  postaction={decorate, decoration={markings, mark=at position 0.53 with {\arrow[fill=black, scale=1.5]{Triangle}} } }] (0.3,0)--(-1.3,0);
    \filldraw [black] (-1,0.05) circle (0pt) node[anchor=south ]{$c^T_{\mu_1} (p) $};
    \filldraw [black] (0,0.05) circle (0pt) node[anchor=south ]{$\tensor{b}{_{\mu_2}} (p) $};
    \filldraw [black] (-0.5,-0.1) circle (0pt) node[anchor=north ]{\large $p$};
    \filldraw [black] (1.7,0) circle (0pt) node[anchor=east ]{$=i\, G^{(c^T_\mu b_\nu)}_{\mu_1\mu_2}(p) $};
\end{tikzpicture}
\end{subfigure}
\begin{subfigure}[b]{0.8 \textwidth}
    \centering
    \begin{tikzpicture}[scale=3,
        decoration={coil,amplitude=4.25,segment length=4.75},anchor=base, baseline]
    \draw[line width=0.8,black,  postaction={decorate, decoration={markings, mark=at position 0.53 with {\arrow[fill=black, scale=1.5]{Triangle}} } }] (0.3,0)--(-1.3,0);
    \filldraw [black] (-1,0.05) circle (0pt) node[anchor=south ]{$c(p) $};
    \filldraw [black] (0,0.05) circle (0pt) node[anchor=south ]{$\tensor{b}{_{\mu_1}} (p) $};
    \filldraw [black] (-0.5,-0.1) circle (0pt) node[anchor=north ]{\large $p$};
    \filldraw [black] (1.7,0) circle (0pt) node[anchor=east ]{$=i\, {G}^{(c b_\nu)}_{\mu_1}(p)$};
\end{tikzpicture}
\end{subfigure}
\begin{subfigure}[b]{0.8 \textwidth}
    \centering
        \begin{tikzpicture}[scale=3,
                decoration={coil,amplitude=4.25,segment length=4.75},anchor=base, baseline]
            \draw[line width=0.8,black,  postaction={decorate, decoration={markings, mark=at position 0.53 with {\arrow[fill=black, scale=1.5]{Triangle}} } }] (0.3,0)--(-1.3,0);
            \filldraw [black] (-1,0.05) circle (0pt) node[anchor=south ]{$c(p) $};
            \filldraw [black] (0,0.05) circle (0pt) node[anchor=south ]{$b(p) $};
            \filldraw [black] (-0.5,-0.1) circle (0pt) node[anchor=north ]{\large $p$};
            \filldraw [black] (1.7,0) circle (0pt) node[anchor=east ]{$=i\, \tensor{G}{^{(c b)}}(p) $};
        \end{tikzpicture}
\end{subfigure}
\caption{Free propagators.} \label{fig:1}
\end{figure}

\begin{figure}[htbp]
    \begin{center}
\begin{subfigure}[b]{0.8 \textwidth}
\centering
    \begin{tikzpicture}[scale=2.4,
        decoration={coil,amplitude=4.25,segment length=4.75},anchor=base, baseline]
        \draw[line width=0.8,black,  postaction={decorate, decoration={markings, mark=at position 1 with {\arrow[fill=black, scale=1.5]{Triangle}} } }] (-0.441,0)--(-0.44,0);
         \draw[spring] (-.5,0) -- (-0.95,0);
         \draw[spring] (-0.45,0) -- (0,0);
         \draw[line width=0.8,black,  postaction={decorate, decoration={markings, mark=at position 1 with {\arrow[fill=black, scale=1.5]{Triangle}} } }] (0.31,0.31)--(0.3,0.3);
         \draw[spring] (.35,0.35) -- (0.6,0.6);
         \draw[spring] (0.305,0.305) -- (0,0);
         \draw[line width=0.8,black,  postaction={decorate, decoration={markings, mark=at position 1 with {\arrow[fill=black, scale=1.5]{Triangle}} } }] (0.31,-0.31)--(0.3,-0.3);
         \draw[spring] (.35,-0.35) -- (0.6,-0.6);
         \draw[spring] (0.305,-0.305) -- (0,0);
       \filldraw [black] (-0.5,-0.1) circle (0pt) node[anchor=north]{\large$p_3$};
       \filldraw [black] (-.5,0.05) circle (0pt) node[anchor=south ]{$\tensor{h}{_{\mu_3}_{\nu_3}} (p_3) $};
       \filldraw [black] (.45,.33) circle (0pt) node[anchor=north ]{\large$p_2$};
       \filldraw [black] (.43,.4) circle (0pt) node[anchor=south east ]{$\tensor{h}{_{\mu_2}_{\nu_2}} (p_2) $};
       \filldraw [black] (.45,-.12) circle (0pt) node[anchor=north ]{\large$p_1$};
       \filldraw [black] (.43,-.65) circle (0pt) node[anchor=south east]{$\tensor{h}{_{\mu_1}_{\nu_1}} (p_1) $};
       \filldraw [black] (3.35,0) circle (0pt) node[anchor= east]{$=V^{\mu_1\nu_1,\mu_2\nu_2,\mu_3\nu_3}(p_1,p_2,p_3) $};
    \end{tikzpicture}
\end{subfigure}
\begin{subfigure}[b]{0.8 \textwidth}
\centering
\begin{tikzpicture}[scale=2.4,
    decoration={coil,amplitude=4.25,segment length=4.75},anchor=base, baseline]
    \draw[line width=0.8,black,  postaction={decorate, decoration={markings, mark=at position 1 with {\arrow[fill=black, scale=1.5]{Triangle}} } }] (-0.441,0)--(-0.44,0);
    \draw[spring] (-.5,0) -- (-0.95,0);
    \draw[spring] (-0.45,0) -- (0,0);
    \draw[line width=0.8,black,  postaction={decorate, decoration={markings, mark=at position 0.3 with {\arrow[fill=black, scale=1.5]{Triangle}} } }] (0.6,0.6)--(0,0);
    \draw[line width=0.8,black,  postaction={decorate, decoration={markings, mark=at position 0.75 with {\arrow[fill=black, scale=1.5]{Triangle}} } }] (0,0) --(0.6,-0.6);
    \fill[black] (0.2,0.2)  -- (0,0)-- (0.2,-0.2) -- cycle;
    \filldraw [black] (-0.5,-0.1) circle (0pt) node[anchor=north]{\large$p_3$};
    \filldraw [black] (-.5,0.05) circle (0pt) node[anchor=south ]{$\tensor{h}{_{\mu_3}_{\nu_3}}(p_3) $};
    \filldraw [black] (.5,.37) circle (0pt) node[anchor=north ]{\large$p_2$};
    \filldraw [black] (.47,.47) circle (0pt) node[anchor=south east ]{$c^T_{\mu_2}(p_2) $};
    \filldraw [black] (.5,-.15) circle (0pt) node[anchor=north ]{\large$p_1$};
    \filldraw [black] (.47,-.67) circle (0pt) node[anchor=south east]{$b_{\mu_1}(p_1) $};
    \filldraw [black] (3,0) circle (0pt) node[anchor= east]{$=V^{\mu_1,\mu_2,\mu_3\nu_3}_1(p_1,p_2,p_3)$};
\end{tikzpicture}
\end{subfigure}
\begin{subfigure}[b]{0.8 \textwidth}
\centering
    \begin{tikzpicture}[scale=2.4,
        decoration={coil,amplitude=4.25,segment length=4.75},anchor=base, baseline]
        \draw[line width=0.8,black,  postaction={decorate, decoration={markings, mark=at position 0.5 with {\arrow[fill=black, scale=1.5]{Triangle}} } }] (0.6,0.6)--(0,0);
        \draw[line width=0.8,black,  postaction={decorate, decoration={markings, mark=at position 0.6 with {\arrow[fill=black, scale=1.5]{Triangle}} } }] (0,0) --(0.6,-0.6);
        \filldraw [black] (0,0) circle (2pt) node[anchor=south]{};
        \draw[line width=0.8,black,  postaction={decorate, decoration={markings, mark=at position 1 with {\arrow[fill=black, scale=1.5]{Triangle}} } }] (-0.441,0)--(-0.44,0);
        \draw[spring] (-.5,0) -- (-0.95,0);
        \draw[spring] (-0.45,0) -- (-.08,0);
        \filldraw [black] (-0.5,-0.1) circle (0pt) node[anchor=north]{\large$p_3$};
        \filldraw [black] (-.5,0.05) circle (0pt) node[anchor=south ]{$\tensor{h}{_{\mu_3}_{\nu_3}}(p_3) $};
        \filldraw [black] (.5,.3) circle (0pt) node[anchor=north ]{\large$p_2$};
        \filldraw [black] (.47,.4) circle (0pt) node[anchor=south east ]{$c(p_2) $};
        \filldraw [black] (.5,-.2) circle (0pt) node[anchor=north ]{\large$p_1$};
        \filldraw [black] (.47,-.65) circle (0pt) node[anchor=south east]{$b_{\mu_1}(p_1) $};
        \filldraw [black] (3,0) circle (0pt) node[anchor= east]{$=V^{\mu_1,\mu_3\nu_3}_2(p_1,p_2,p_3)$};
    \end{tikzpicture}
\end{subfigure}
\begin{subfigure}[b]{0.8 \textwidth}
\centering
    \begin{tikzpicture}[scale=2.4,
        decoration={coil,amplitude=4.25,segment length=4.75},anchor=base, baseline]
        \draw[line width=0.8,black,  postaction={decorate, decoration={markings, mark=at position 1 with {\arrow[fill=black, scale=1.5]{Triangle}} } }] (-0.441,0)--(-0.44,0);
        \draw[spring] (-.5,0) -- (-0.95,0);
        \draw[spring] (-0.45,0) -- (-.12,0);
        \draw[line width=0.8,black,  postaction={decorate, decoration={markings, mark=at position 0.5 with {\arrow[fill=black, scale=1.5]{Triangle}} } }] (0.6,0.6)--(0,0);
        \draw[line width=0.8,black,  postaction={decorate, decoration={markings, mark=at position 0.6 with {\arrow[fill=black, scale=1.5]{Triangle}} } }] (0,0) --(0.6,-0.6);
        \fill[black] (0.1,0.1) arc[start angle=45, end angle=315, radius=0.1414] --(0,0)-- cycle;
        \filldraw [black] (-0.5,-0.1) circle (0pt) node[anchor=north]{\large$p_3$};
        \filldraw [black] (-.5,0.05) circle (0pt) node[anchor=south ]{$\tensor{h}{_{\mu_3}_{\nu_3}}(p_3) $};
        \filldraw [black] (.5,.3) circle (0pt) node[anchor=north ]{\large$p_2$};
        \filldraw [black] (.47,.4) circle (0pt) node[anchor=south east ]{$c(p_2) $};
        \filldraw [black] (.5,-.2) circle (0pt) node[anchor=north ]{\large$p_1$};
        \filldraw [black] (.47,-.65) circle (0pt) node[anchor=south east]{$b(p_1) $};
        \filldraw [black] (3,0) circle (0pt) node[anchor= east]{$=V^{\mu_3\nu_3}_3(p_1,p_2,p_3)$};
    \end{tikzpicture}

\end{subfigure}
\begin{subfigure}[b]{0.8 \textwidth}
\centering
    \begin{tikzpicture}[scale=2.4,
        decoration={coil,amplitude=4.25,segment length=4.75},anchor=base, baseline]
        \draw[line width=0.8,black,  postaction={decorate, decoration={markings, mark=at position 1 with {\arrow[fill=black, scale=1.5]{Triangle}} } }] (-0.441,0)--(-0.44,0);
        \draw[spring] (-.5,0) -- (-0.95,0);
        \draw[spring] (-0.45,0) -- (0,0);
        \draw[line width=0.8,black,  postaction={decorate, decoration={markings, mark=at position 0.5 with {\arrow[fill=black, scale=1.5]{Triangle}} } }] (0.6,0.6)--(0,0);
        \draw[line width=0.8,black,  postaction={decorate, decoration={markings, mark=at position 0.6 with {\arrow[fill=black, scale=1.5]{Triangle}} } }] (0,0) --(0.6,-0.6);
        \filldraw [black] (-0.5,-0.1) circle (0pt) node[anchor=north]{\large$p_3$};
        \filldraw [black] (-.5,0.05) circle (0pt) node[anchor=south ]{$\tensor{h}{_{\mu_3}_{\nu_3}}(p_3) $};
        \filldraw [black] (.5,.3) circle (0pt) node[anchor=north ]{\large$p_2$};
        \filldraw [black] (.47,.4) circle (0pt) node[anchor=south east ]{$c^T_{\mu_2}(p_2) $};
        \filldraw [black] (.5,-.2) circle (0pt) node[anchor=north ]{\large$p_1$};
        \filldraw [black] (.47,-.65) circle (0pt) node[anchor=south east]{$b(p_1) $};
        \filldraw [black] (3,0) circle (0pt) node[anchor= east]{$=V^{\mu_2,\mu_3\nu_3}_4(p_1,p_2,p_3)$};
        \end{tikzpicture}
\end{subfigure}
\begin{subfigure}[b]{0.8 \textwidth}
\centering
    \begin{tikzpicture}[scale=2.4,
        decoration={coil,amplitude=4.25,segment length=4.75},anchor=base, baseline]
        \draw[line width=0.8,black] (-0.75,0) -- (0,0);
        \draw[line width=0.8,black,  postaction={decorate, decoration={markings, mark=at position 0.5 with {\arrow[fill=black, scale=1.5]{Triangle}} } }] (0.6,0.6)--(0,0);
        \draw[line width=0.8,black,  postaction={decorate, decoration={markings, mark=at position 0.6 with {\arrow[fill=black, scale=1.5]{Triangle}} } }] (0,0) --(0.6,-0.6);
        \filldraw [black] (-0.5,-0.1) circle (0pt) node[anchor=north]{\large$p_3$};
        \filldraw [black] (-.5,0.05) circle (0pt) node[anchor=south ]{$f(p_3) $};
        \filldraw [black] (.5,.3) circle (0pt) node[anchor=north ]{\large$p_2$};
        \filldraw [black] (.47,.4) circle (0pt) node[anchor=south east ]{$c^T_{{\mu_2}}(p_2) $};
        \filldraw [black] (.5,-.2) circle (0pt) node[anchor=north ]{\large$p_1$};
        \filldraw [black] (.47,-.65) circle (0pt) node[anchor=south east]{$b(p_1) $};
        \filldraw [black] (3,0) circle (0pt) node[anchor= east]{$=V^{\mu_2}_5(p_1,p_2,p_3)$};

    \end{tikzpicture}
\end{subfigure}
\end{center}
\caption{Three-field Vertices.} \label{fig:2}
\end{figure}
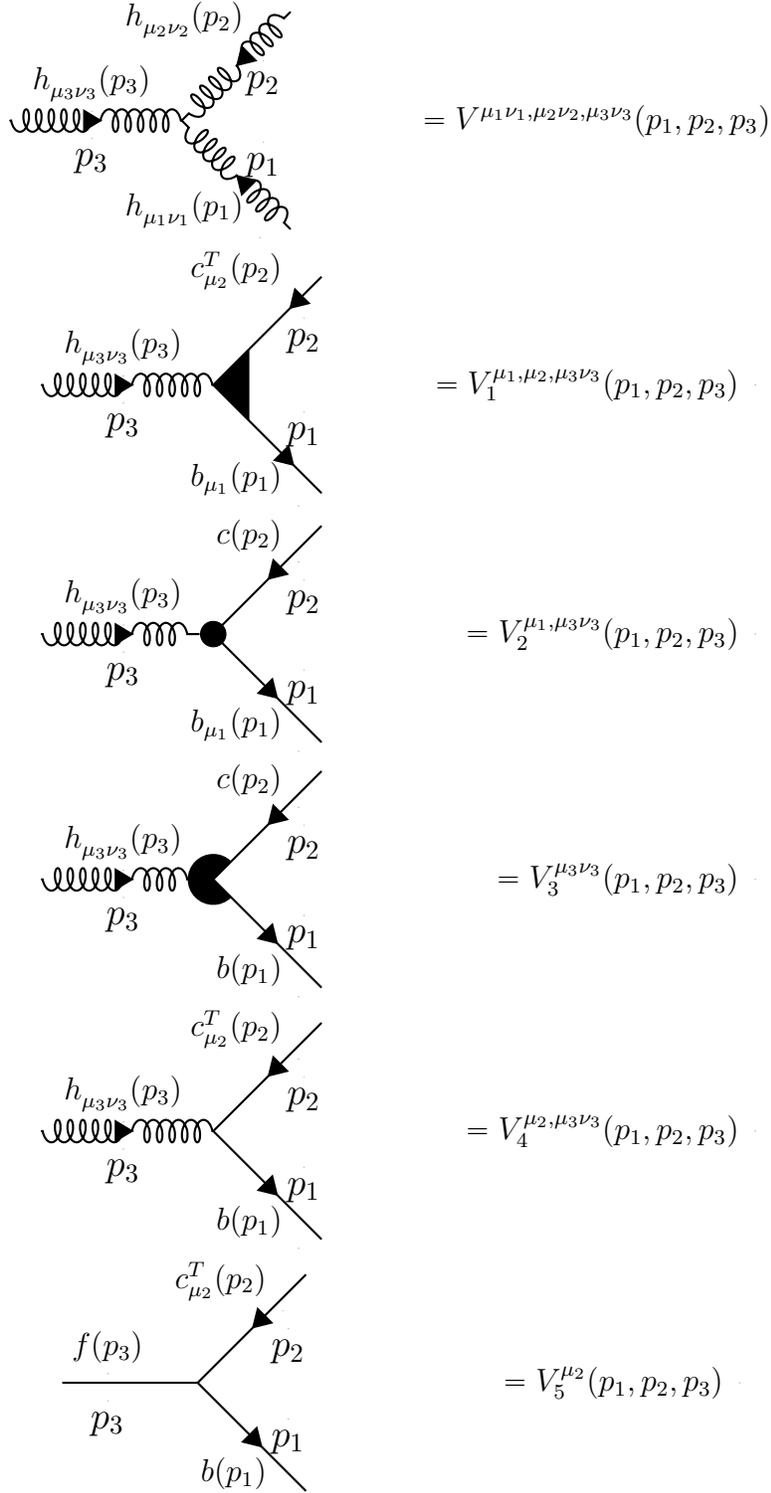

\section{The one-loop contribution to the graviton propagator.}

Let $iG^{(hh\,(1))}_{\m_1\m_2\m_3\m_4}(p)$ denote the one-loop contribution to the graviton propagator. Then, we have
{
\begin{equation}
\begin{array}{l}
{iG^{(hh\,(1))}_{\m_1\m_2\m_3\m_4}(p)=iG^{(hh)}_{\m_1\m_2\r_1\r_2}(p)\,\Gamma^{(hh)}_{\r_1\r_2\r_3\r_4}(p)\,iG^{(hh)}_{\r_3\r_4\m_3\m_4}(p)+}\\[4pt]
{\phantom{iG^{(hh\,(1))}_{\m_1\m_2\m_3\m_4}(p)=}
iG^{(hf)}_{\m_1\m_2}(p)\,\Gamma^{(fh)}_{\r_3\r_4}(p)\,iG^{(hh)}_{\r_3\r_4\m_3\m_4}(p)+
iG^{(hh)}_{\m_1\m_2\r_1\r_2}(p)\,\Gamma^{(hf)}_{\r_1\r_2}(p)\,iG^{(hf)}_{\m_3\m_4}(p).}
\end{array}
\label{oneloopprop}
\end{equation}
}
where $G^{(hh)}_{\m_1\m_2\m_3\m_4}(p)$ and $G^{(hf)}_{\m_1\m_2}(p)$ have been given in (\ref{Gpropaone}) and (\ref{Gpropatwo}), respectively.  $\Gamma^{(hh)}_{\r_1\r_2\r_3\r_4}(p)$ is the sum of the values of all the 1PI diagrams in {\cref{fig:3}},  $\Gamma^{(fh)}_{\r_3\r_4}(p)$ is given the value of the 1PI diagram referred to as Diagram 8 in {\cref{fig:4}}. Also in {\cref{fig:4}}, the 1PI diagram called Diagram 9 yields $\Gamma^{(hf)}_{\r_1\r_2}(p)$. The value of each 1PI Feynman diagram in {\cref{fig:3}} can be found in {Appendix} B. In the current section, we shall display the value of $\Gamma^{(hh)}_{\r_1\r_2\r_3\r_4}(p)$, $\Gamma^{(hf)}_{\r_1\r_2}(p)$ and $\Gamma^{(fh)}_{\r_1\r_2}(p)$ that, for arbitrary complex values of $D$, we have obtained by using dimensional regularization.

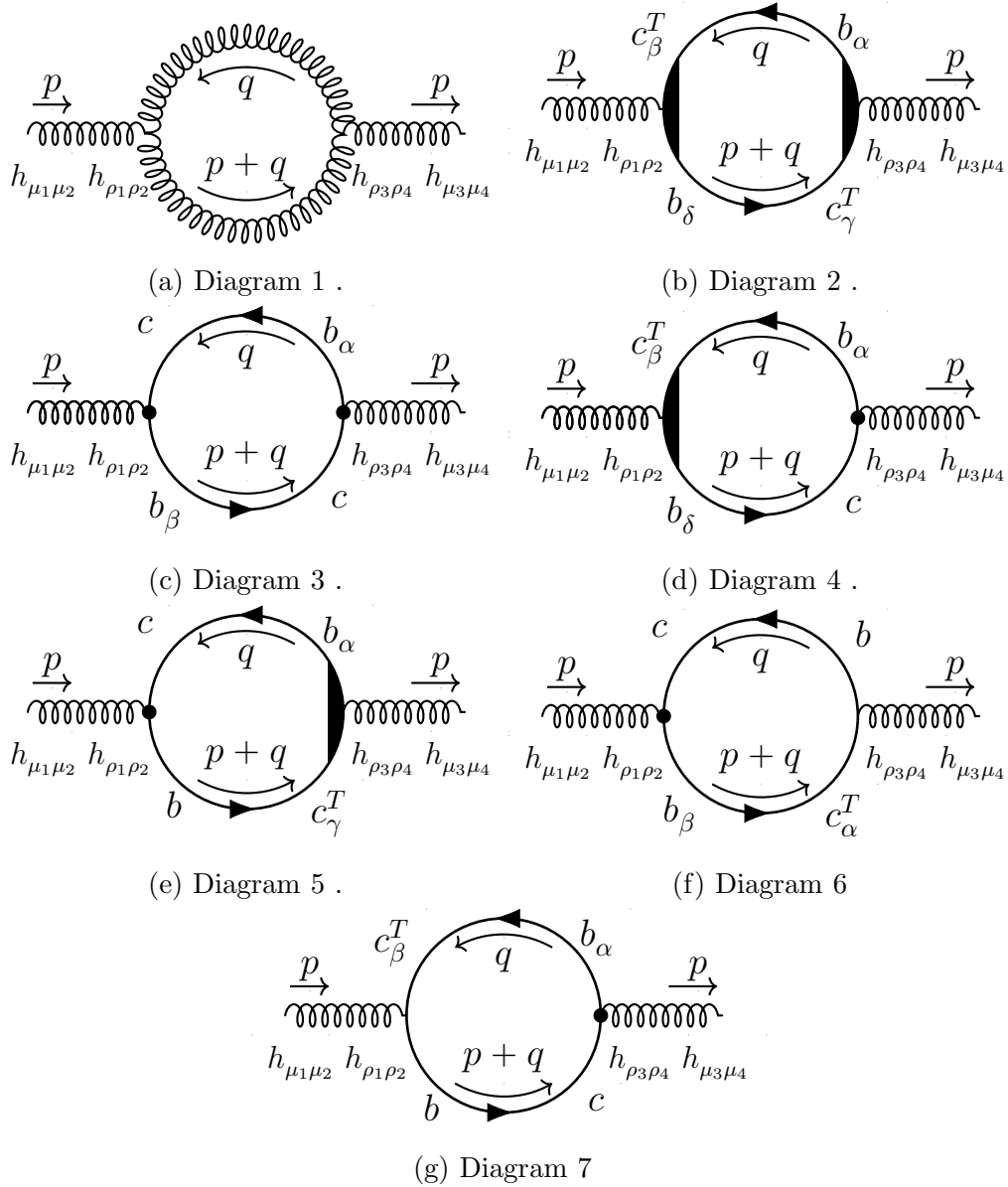
\begin{figure}[htbp]
    \centering
\begin{subfigure}[b]{0.4 \textwidth}
    \begin{tikzpicture}[scale=1.3,
        decoration={coil,amplitude=4.25,segment length=4.75},anchor=base, baseline]
        \draw[decorate,line width=0.8,black] (0,0) arc (0:180:1);
        \draw[decorate,line width=0.8,black] (-2,0) arc (-180:0:1);
        \draw[spring] (-3.25,0) -- (-2,0);
        \draw[spring] (0,0) -- (1.25,0);
    \draw[thick, <-,black](1.2,0.3) -- (0.7,0.3);
    \draw[thick, ->,black](-0.5,0.55) arc (60:120:1);
    \draw[thick, <-,black](-0.5,-0.55) arc (-60:-120:1);
    \filldraw [black] (-1,0.7) circle (0pt) node[anchor=north]{\large$q$};
    \filldraw [black] (-1,-0.6) circle (0pt) node[anchor=south]{\large$p+q$};
    \filldraw [black] (-3.6,-0.2) circle (0pt) node[anchor=north west]{$\tensor{h}{_{\mu_1}_{\mu_2}} $};
    \filldraw [black] (-2.8,-0.2) circle (0pt) node[anchor=north west]{$\tensor{h}{_{\rho_1}_{\rho_2}} $};
    \filldraw [black] (1.7,-0.2) circle (0pt) node[anchor=north east]{$\tensor{h}{_{\mu_3}_{\mu_4}} $};
    \filldraw [black] (.9,-0.2) circle (0pt) node[anchor=north east]{$\tensor{h}{_{\rho_3}_{\rho_4}} $};
    \filldraw [black] (-3,0.25) circle (0pt) node[anchor=south]{\large$p$};
    \draw[thick, ->,black](-3.2,0.3) -- (-2.8,0.3);
    \filldraw [black] (1,0.25) circle (0pt) node[anchor=south]{\large$p$};

        \end{tikzpicture}
           \caption{Diagram 1 .}\label{fig:DG1}
    \end{subfigure}
    \begin{subfigure}[b]{0.4\textwidth}
    \begin{tikzpicture}[scale=1.3,
        decoration={coil,amplitude=4.25,segment length=4.75},anchor=base, baseline]
        \filldraw [black] (-3,0.25) circle (0pt) node[anchor=south]{\large$p$};
        \draw[thick, ->,black](-3.2,0.3) -- (-2.8,0.3);
        \filldraw [black] (1,0.25) circle (0pt) node[anchor=south]{\large$p$};
        \draw[line width=1] (0,0) arc (0:180:1);
        \draw[   -{Latex[length=3.5mm]}] (-1.09,1)--(-1.1,1);
        \draw[line width=1] (-2,0) arc (-180:0:1);
         \draw[   -{Latex[length=3.5mm]}] (-.95,-1)--(-.9,-1);
        \draw[spring] (0,0) -- (1.25,0);
       \draw[line width=0.8,black] (-1.85,0.5268)--(-1.85,-0.5268);
       \draw[line width=0.8,black] (-0.15,0.5268)--(-0.15,-0.5268);
       \fill[black] (-1.85,-0.5268)-- (-1.85,0.5268) arc[start angle=148.212, end angle=211.788, radius=1] -- cycle;
       \fill[black] (-0.15,-0.5268)-- (-0.15,0.5268) arc[start angle=31.788, end angle=-31.788, radius=1] -- cycle;
     \draw[thick, <-,black](1.2,0.3) -- (0.7,0.3);
     \draw[thick, ->,black](-0.5,0.7) arc (60:120:1);
     \draw[thick, <-,black](-0.5,-0.7) arc (-60:-120:1);
     \draw[spring] (-3.25,0) -- (-2,0);
     \filldraw [black] (-1,0.8) circle (0pt) node[anchor=north]{\large$q$};
     \filldraw [black] (-1,-0.7) circle (0pt) node[anchor=south]{\large$p+q$};
     \filldraw [black] (-3.6,-0.2) circle (0pt) node[anchor=north west]{$\tensor{h}{_{\mu_1}_{\mu_2}} $};
     \filldraw [black] (-2.8,-0.2) circle (0pt) node[anchor=north west]{$\tensor{h}{_{\rho_1}_{\rho_2}} $};
     \filldraw [black] (1.7,-0.2) circle (0pt) node[anchor=north east]{$\tensor{h}{_{\mu_3}_{\mu_4}} $};
     \filldraw [black] (.9,-0.2) circle (0pt) node[anchor=north east]{$\tensor{h}{_{\rho_3}_{\rho_4}} $};
     \filldraw [black] (0.3,1.1) circle (0pt) node[anchor=north east]{\large$\tensor{b}{_\alpha} $};
     \filldraw [black] (-1.8,1.1) circle (0pt) node[anchor=north east]{\large$c^T_\beta $};
     \filldraw [black] (0.2,-0.7) circle (0pt) node[anchor=north east]{\large$c^T_\gamma $};
     \filldraw [black] (-1.5,-0.7) circle (0pt) node[anchor=north east]{\large$\tensor{b}{_\delta} $};

    \end{tikzpicture}
    \caption{Diagram 2 .}\label{fig:DG2}
\end{subfigure}

    \begin{subfigure}[b]{0.4\textwidth}
        \begin{tikzpicture}[scale=1.3,
            decoration={coil,amplitude=4.25,segment length=4.75},anchor=base, baseline]

           \draw[spring] (-3.25,0) -- (-2,0);
           \draw[line width=1] (0,0) arc (0:180:1);
            \draw[   -{Latex[length=3.5mm]}] (-1.09,1)--(-1.1,1);
            \draw[line width=1] (-2,0) arc (-180:0:1);
             \draw[   -{Latex[length=3.5mm]}] (-.95,-1)--(-.9,-1);
            \draw[spring] (0,0) -- (1.25,0);
            \filldraw [black] (-2,0) circle (2pt) node[anchor=south]{};
            \filldraw [black] (0,0) circle (2pt) node[anchor=south]{};

        \draw[thick, <-,black](1.2,0.3) -- (0.7,0.3);
        \draw[thick, ->,black](-0.5,0.7) arc (60:120:1);
        \draw[thick, <-,black](-0.5,-0.7) arc (-60:-120:1);
        \draw[spring] (-3.25,0) -- (-2,0);
        \filldraw [black] (-1,0.8) circle (0pt) node[anchor=north]{\large$q$};
        \filldraw [black] (-1,-0.7) circle (0pt) node[anchor=south]{\large$p+q$};
        \filldraw [black] (-3.6,-0.2) circle (0pt) node[anchor=north west]{$\tensor{h}{_{\mu_1}_{\mu_2}} $};
        \filldraw [black] (-2.8,-0.2) circle (0pt) node[anchor=north west]{$\tensor{h}{_{\rho_1}_{\rho_2}} $};
        \filldraw [black] (1.7,-0.2) circle (0pt) node[anchor=north east]{$\tensor{h}{_{\mu_3}_{\mu_4}} $};
        \filldraw [black] (.9,-0.2) circle (0pt) node[anchor=north east]{$\tensor{h}{_{\rho_3}_{\rho_4}} $};
        \filldraw [black] (0.3,1.1) circle (0pt) node[anchor=north east]{\large$\tensor{b}{_\alpha} $};
        \filldraw [black] (-1.8,1.1) circle (0pt) node[anchor=north east]{\large$c $};
        \filldraw [black] (0.2,-0.7) circle (0pt) node[anchor=north east]{\large$c $};
        \filldraw [black] (-1.5,-0.7) circle (0pt) node[anchor=north east]{\large$\tensor{b}{_\beta} $};
        \filldraw [black] (-3,0.25) circle (0pt) node[anchor=south]{\large$p$};
        \draw[thick, ->,black](-3.2,0.3) -- (-2.8,0.3);
        \filldraw [black] (1,0.25) circle (0pt) node[anchor=south]{\large$p$};

            \end{tikzpicture}
           \caption{Diagram 3 .}\label{fig:DG3}
\end{subfigure}
\begin{subfigure}[b]{0.4\textwidth}

    \begin{tikzpicture}[scale=1.3,
     decoration={coil,amplitude=4.25,segment length=4.75},anchor=base, baseline]
     \draw[spring] (-3.25,0) -- (-2,0);
           \draw[line width=1] (0,0) arc (0:180:1);
            \draw[   -{Latex[length=3.5mm]}] (-1.09,1)--(-1.1,1);
            \draw[line width=1] (-2,0) arc (-180:0:1);
             \draw[   -{Latex[length=3.5mm]}] (-.95,-1)--(-.9,-1);
     \draw[spring] (0,0) -- (1.25,0);

     \filldraw [black] (0,0) circle (2pt) node[anchor=south]{};
     \draw[line width=0.8,black] (-1.85,0.5268)--(-1.85,-0.5268);
     \fill[black] (-1.85,-0.5268)-- (-1.85,0.5268) arc[start angle=148.212, end angle=211.788, radius=1] -- cycle;

    \draw[thick, <-,black](1.2,0.3) -- (0.7,0.3);
    \draw[thick, ->,black](-0.5,0.7) arc (60:120:1);
    \draw[thick, <-,black](-0.5,-0.7) arc (-60:-120:1);
    \draw[spring] (-3.25,0) -- (-2,0);
    \filldraw [black] (-1,0.8) circle (0pt) node[anchor=north]{\large$q$};
    \filldraw [black] (-1,-0.7) circle (0pt) node[anchor=south]{\large$p+q$};
    \filldraw [black] (-3.6,-0.2) circle (0pt) node[anchor=north west]{$\tensor{h}{_{\mu_1}_{\mu_2}} $};
    \filldraw [black] (-2.8,-0.2) circle (0pt) node[anchor=north west]{$\tensor{h}{_{\rho_1}_{\rho_2}} $};
    \filldraw [black] (1.7,-0.2) circle (0pt) node[anchor=north east]{$\tensor{h}{_{\mu_3}_{\mu_4}} $};
    \filldraw [black] (.9,-0.2) circle (0pt) node[anchor=north east]{$\tensor{h}{_{\rho_3}_{\rho_4}} $};
    \filldraw [black] (0.3,1.1) circle (0pt) node[anchor=north east]{\large$\tensor{b}{_\alpha} $};
    \filldraw [black] (-1.8,1.1) circle (0pt) node[anchor=north east]{\large$c^T_\beta $};
    \filldraw [black] (0.2,-0.7) circle (0pt) node[anchor=north east]{\large$c $};
    \filldraw [black] (-1.5,-0.7) circle (0pt) node[anchor=north east]{\large$\tensor{b}{_\delta} $};
    \filldraw [black] (-3,0.25) circle (0pt) node[anchor=south]{\large$p$};
    \draw[thick, ->,black](-3.2,0.3) -- (-2.8,0.3);
    \filldraw [black] (1,0.25) circle (0pt) node[anchor=south]{\large$p$};
    \end{tikzpicture}
    \caption{Diagram 4  .}\label{fig:DG4}
\end{subfigure}
\begin{subfigure}[b]{0.4\textwidth}
    \begin{tikzpicture}[scale=1.3,
        decoration={coil,amplitude=4.25,segment length=4.75},anchor=base, baseline]
             \draw[line width=1] (0,0) arc (0:180:1);
            \draw[   -{Latex[length=3.5mm]}] (-1.09,1)--(-1.1,1);
            \draw[line width=1] (-2,0) arc (-180:0:1);
             \draw[   -{Latex[length=3.5mm]}] (-.95,-1)--(-.9,-1);
        \draw[spring] (0,0) -- (1.25,0);

        \filldraw [black] (-2,0) circle (2pt) node[anchor=south]{};
        \draw[line width=0.8,black] (-0.15,0.5268)--(-0.15,-0.5268);
        \fill[black] (-0.15,-0.5268)-- (-0.15,0.5268) arc[start angle=31.788, end angle=-31.788, radius=1] -- cycle;
    \draw[thick, <-,black](1.2,0.3) -- (0.7,0.3);
    \draw[thick, ->,black](-0.5,0.7) arc (60:120:1);
    \draw[thick, <-,black](-0.5,-0.7) arc (-60:-120:1);
    \draw[spring] (-3.25,0) -- (-2,0);
    \filldraw [black] (-1,0.8) circle (0pt) node[anchor=north]{\large$q$};
    \filldraw [black] (-1,-0.7) circle (0pt) node[anchor=south]{\large$p+q$};
    \filldraw [black] (-3.6,-0.2) circle (0pt) node[anchor=north west]{$\tensor{h}{_{\mu_1}_{\mu_2}} $};
    \filldraw [black] (-2.8,-0.2) circle (0pt) node[anchor=north west]{$\tensor{h}{_{\rho_1}_{\rho_2}} $};
    \filldraw [black] (1.7,-0.2) circle (0pt) node[anchor=north east]{$\tensor{h}{_{\mu_3}_{\mu_4}} $};
    \filldraw [black] (.9,-0.2) circle (0pt) node[anchor=north east]{$\tensor{h}{_{\rho_3}_{\rho_4}} $};
    \filldraw [black] (0.3,1.1) circle (0pt) node[anchor=north east]{\large$\tensor{b}{_\alpha} $};
    \filldraw [black] (-1.8,1.1) circle (0pt) node[anchor=north east]{\large$c$};
    \filldraw [black] (0.2,-0.7) circle (0pt) node[anchor=north east]{\large$c^T_\gamma $};
    \filldraw [black] (-1.5,-0.7) circle (0pt) node[anchor=north east]{\large$b $};
    \filldraw [black] (-3,0.25) circle (0pt) node[anchor=south]{\large$p$};
    \draw[thick, ->,black](-3.2,0.3) -- (-2.8,0.3);
    \filldraw [black] (1,0.25) circle (0pt) node[anchor=south]{\large$p$};

    \end{tikzpicture}
       \caption{Diagram 5  .}\label{fig:DG5}
       \end{subfigure}
       \begin{subfigure}[b]{0.4\textwidth}
        \begin{tikzpicture}[scale=1.3,
            decoration={coil,amplitude=4.25,segment length=4.75},anchor=base, baseline]
                 \draw[line width=1] (0,0) arc (0:180:1);
    \draw[   -{Latex[length=3.5mm]}] (-1.09,1)--(-1.1,1);
    \draw[line width=1] (-2,0) arc (-180:0:1);
     \draw[   -{Latex[length=3.5mm]}] (-.95,-1)--(-.9,-1);
            \draw[spring] (0,0) -- (1.25,0);
            \filldraw [black] (-2,0) circle (2pt) node[anchor=south]{};
\draw[thick, <-,black](1.2,0.3) -- (0.7,0.3);
\draw[thick, ->,black](-0.5,0.7) arc (60:120:1);
\draw[thick, <-,black](-0.5,-0.7) arc (-60:-120:1);
\draw[spring] (-3.25,0) -- (-2,0);
\filldraw [black] (-1,0.8) circle (0pt) node[anchor=north]{\large$q$};
\filldraw [black] (-1,-0.7) circle (0pt) node[anchor=south]{\large$p+q$};
\filldraw [black] (-3.6,-0.2) circle (0pt) node[anchor=north west]{$\tensor{h}{_{\mu_1}_{\mu_2}} $};
\filldraw [black] (-2.8,-0.2) circle (0pt) node[anchor=north west]{$\tensor{h}{_{\rho_1}_{\rho_2}} $};
\filldraw [black] (1.7,-0.2) circle (0pt) node[anchor=north east]{$\tensor{h}{_{\mu_3}_{\mu_4}} $};
\filldraw [black] (.9,-0.2) circle (0pt) node[anchor=north east]{$\tensor{h}{_{\rho_3}_{\rho_4}} $};
\filldraw [black] (0.3,1.1) circle (0pt) node[anchor=north east]{\large$b $};
\filldraw [black] (-1.8,1.1) circle (0pt) node[anchor=north east]{\large$c $};
\filldraw [black] (0.2,-0.7) circle (0pt) node[anchor=north east]{\large$c^T_\alpha $};
\filldraw [black] (-1.5,-0.7) circle (0pt) node[anchor=north east]{\large$\tensor{b}{_\beta} $};
\filldraw [black] (-3,0.25) circle (0pt) node[anchor=south]{\large$p$};
\draw[thick, ->,black](-3.2,0.3) -- (-2.8,0.3);
\filldraw [black] (1,0.25) circle (0pt) node[anchor=south]{\large$p$};
        \end{tikzpicture}
        \caption{Diagram 6  }\label{fig:DG6}
    \end{subfigure}
\begin{subfigure}[b]{0.4\textwidth}
\begin{tikzpicture}[scale=1.3,
    decoration={coil,amplitude=4.25,segment length=4.75},anchor=base, baseline]
    \draw[line width=1] (0,0) arc (0:180:1);
    \draw[   -{Latex[length=3.5mm]}] (-1.09,1)--(-1.1,1);
    \draw[line width=1] (-2,0) arc (-180:0:1);
     \draw[   -{Latex[length=3.5mm]}] (-.95,-1)--(-.9,-1);
    \draw[spring] (0,0) -- (1.25,0);
    \filldraw [black] (0,0) circle (2pt) node[anchor=south]{};
\draw[thick, <-,black](1.2,0.3) -- (0.7,0.3);
\draw[thick, ->,black](-0.5,0.7) arc (60:120:1);
\draw[thick, <-,black](-0.5,-0.7) arc (-60:-120:1);
\draw[spring] (-3.25,0) -- (-2,0);
\filldraw [black] (-1,0.8) circle (0pt) node[anchor=north]{\large$q$};
\filldraw [black] (-1,-0.7) circle (0pt) node[anchor=south]{\large$p+q$};
\filldraw [black] (-3.6,-0.2) circle (0pt) node[anchor=north west]{$\tensor{h}{_{\mu_1}_{\mu_2}} $};
\filldraw [black] (-2.8,-0.2) circle (0pt) node[anchor=north west]{$\tensor{h}{_{\rho_1}_{\rho_2}} $};
\filldraw [black] (1.7,-0.2) circle (0pt) node[anchor=north east]{$\tensor{h}{_{\mu_3}_{\mu_4}} $};
\filldraw [black] (.9,-0.2) circle (0pt) node[anchor=north east]{$\tensor{h}{_{\rho_3}_{\rho_4}} $};
\filldraw [black] (0.3,1.1) circle (0pt) node[anchor=north east]{\large$\tensor{b}{_\alpha} $};
\filldraw [black] (-1.8,1.1) circle (0pt) node[anchor=north east]{\large$c^T_\beta $};
\filldraw [black] (0.2,-0.7) circle (0pt) node[anchor=north east]{\large$c $};
\filldraw [black] (-1.5,-0.7) circle (0pt) node[anchor=north east]{\large$b$};
\filldraw [black] (-3,0.25) circle (0pt) node[anchor=south]{\large$p$};
\draw[thick, ->,black](-3.2,0.3) -- (-2.8,0.3);
\filldraw [black] (1,0.25) circle (0pt) node[anchor=south]{\large$p$};
\end{tikzpicture}
\caption{Diagram 7}\label{fig:DG7}
\end{subfigure}

\caption{Non-vanishing contributions to $\Gamma^{(h h)}_{\rho_1 \rho_2 \rho_3 \rho_4}(p)$. }\label{fig:3}
    \end{figure}

\newpage
            \begin{figure}
                \begin{centering}
                \begin{subfigure}[b]{0.8\textwidth}
                    \centering
                    \begin{tikzpicture}[scale=1.7,
                        decoration={coil,amplitude=4.25,segment length=4.75},anchor=base, baseline]
                             \draw[line width=1] (0,0) arc (0:180:1);
            \draw[   -{Latex[length=3.5mm]}] (-1.09,1)--(-1.1,1);
            \draw[line width=1] (-2,0) arc (-180:0:1);
             \draw[   -{Latex[length=3.5mm]}] (-.95,-1)--(-.9,-1);
                        \draw[spring] (0,0) -- (2.2,0);
                        \filldraw [black] (0,0) circle (2pt) node[anchor=south]{};
                    \draw[thick, <-,black](1.2,0.3) -- (0.7,0.3);
                    \draw[thick, ->,black](-0.5,0.7) arc (60:120:1);
                    \draw[thick, <-,black](-0.5,-0.7) arc (-60:-120:1);
                    \draw[spring] (-4.2,0) -- (-3.1,0);
                    \draw[black,thick] (-3.1,0)--(-2,0);
                    \filldraw [black] (-1,0.8) circle (0pt) node[anchor=north]{\large$q$};
                    \filldraw [black] (-1,-0.7) circle (0pt) node[anchor=south]{\large$p+q$};
                    \filldraw [black] (-3.2,-0.15) circle (0pt) node[anchor=north east]{\large$\tensor{h}{_{\mu_1}_{\mu_2}} $};
                    \filldraw [black] (1.5,-0.15) circle (0pt) node[anchor=north west]{\large$\tensor{h}{_{\mu_3}_{\mu_4} }$};
                    \filldraw [black] (0.3,1.1) circle (0pt) node[anchor=north east]{\large$\tensor{b}{_\alpha} $};
                    \filldraw [black] (-1.8,1.1) circle (0pt) node[anchor=north east]{\large$c^T_\beta $};
                    \filldraw [black] (-3,-0.15) circle (0pt) node[anchor=north west]{\large$f $};
                    \filldraw [black] (-0.15,-0.7) circle (0pt) node[anchor=north]{\large$c$};
                    \filldraw [black] (-1.8,-0.7) circle (0pt) node[anchor=north ]{\large$b$};
                    \filldraw [black] (-3,0.25) circle (0pt) node[anchor=south]{\large$p$};
                    \filldraw [black] (.8,-0.15) circle (0pt) node[anchor=north east]{\large$\tensor{h}{_{\rho_3}_{\rho_4}} $};
                    \draw[thick, ->,black](-3.2,0.3) -- (-2.8,0.3);
                    \filldraw [black] (1,0.25) circle (0pt) node[anchor=south]{\large$p$};
                    \end{tikzpicture}
                    \caption{Diagram 8 }\label{fig:DG9}
                \end{subfigure}\vspace{.5 cm}
                \begin{subfigure}[b]{0.8\textwidth}
                    \centering
                    \begin{tikzpicture}[scale=1.7,
                        decoration={coil,amplitude=4.25,segment length=4.75},anchor=base, baseline]
                            \draw[line width=1] (0,0) arc (0:180:1);
           		 \draw[   -{Latex[length=3.5mm]}] (-1.09,1)--(-1.1,1);
           		 \draw[line width=1] (-2,0) arc (-180:0:1);
           		  \draw[   -{Latex[length=3.5mm]}] (-.95,-1)--(-.9,-1);
                        \draw[thick, black] (0,0) -- (1.1,0);
                        \draw[spring, black] (1.1,0) -- (2.2,0);
                        \filldraw [black] (-2,0) circle (2pt) node[anchor=south]{};
            \draw[thick, <-,black](1.2,0.3) -- (0.7,0.3);
            \draw[thick, ->,black](-0.5,0.7) arc (60:120:1);
            \draw[thick, <-,black](-0.5,-0.7) arc (-60:-120:1);
            \draw[spring] (-4.2,0) -- (-2,0);
            \filldraw [black] (-1,0.8) circle (0pt) node[anchor=north]{\large$q$};
            \filldraw [black] (-1,-0.7) circle (0pt) node[anchor=south]{\large$p+q$};
            \filldraw [black] (-3.2,-0.15) circle (0pt) node[anchor=north east]{\large$\tensor{h}{_{\mu_1}_{\mu_2}} $};
            \filldraw [black] (1.5,-0.15) circle (0pt) node[anchor=north west]{\large$\tensor{h}{_{\mu_3}_{\mu_4}} $};
            \filldraw [black] (-2.,-0.15) circle (0pt) node[anchor=north east]{\large$\tensor{h}{_{\rho_1}_{\rho_2}} $};
            \filldraw [black] (.6,-0.15) circle (0pt) node[anchor=north ]{\large$f $};
            \filldraw [black] (0.3,1.1) circle (0pt) node[anchor=north east]{\large$b $};
            \filldraw [black] (-1.8,1.1) circle (0pt) node[anchor=north east]{\large$c $};
            \filldraw [black] (0.1,-0.8) circle (0pt) node[anchor=north east]{\large$c^T_\alpha $};
            \filldraw [black] (-1.5,-0.8) circle (0pt) node[anchor=north east]{\large$\tensor{b}{_\beta} $};
            \filldraw [black] (-3,0.25) circle (0pt) node[anchor=south]{\large$p$};
            \draw[thick, ->,black](-3.2,0.3) -- (-2.8,0.3);
            \filldraw [black] (1,0.25) circle (0pt) node[anchor=south]{\large$p$};
                    \end{tikzpicture}

                    \caption{Diagram 9}\label{fig:DG8}
                \end{subfigure}

                \caption{Non-vanishing contributions to $\Gamma^{(f h)}_{\rho_3 \rho_4 }(p)$ and $\Gamma^{(h f)}_{\rho_1 \rho_2 }(p)$. }\label{fig:3}.
                    \label{fig:4}
            \end{centering}
                    \end{figure}
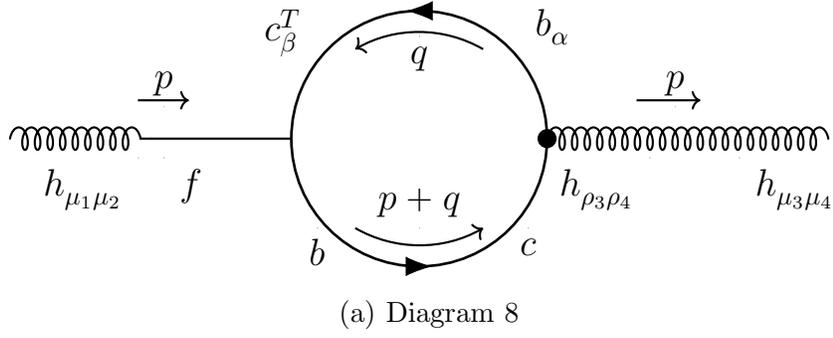
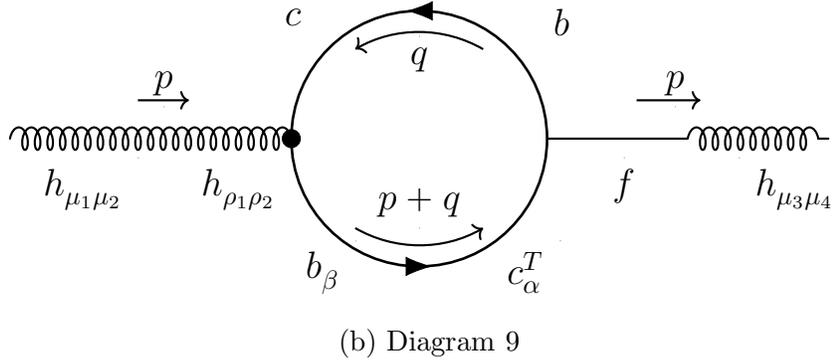

A very involved computation --a computation not feasible without {the} invaluable help from FORM-- leads to the following result
\begin{equation}
\begin{array}{l}
{\Gamma^{(hh)}_{\r_1\r_2\r_3\r_4}(p)=}\\[2pt]
{P_1(p)\,(p^2)^2\left(\eta_{\r_1\r_3}\eta_{\r_2\r_4}+\eta_{\r_1\r_4}\eta_{\r_2\r_3}\right)
+P_2(p)\,(p^2)^2\eta_{\r_1\r_2}\eta_{\r_3\r_4}
+P_3(p)\,p^2\left(\eta_{\r_3\r_4}p_{\r_1} p_{\r_2}+\eta_{\r_1\r_2}p_{\r_3} p_{\r_4}\right)+}\\[4pt]
{P_4(p)\,p^2\left(\eta_{\r_1\r_3}p_{\r_2} p_{\r_4}+\eta_{\r_1\r_4}p_{\r_2} p_{\r_3}+\eta_{\r_2\r_3}p_{\r_1} p_{\r_4}+\eta_{\r_2\r_4}p_{\r_1} p_{\r_3}\right)+P_5(p)\, p_{\r_1} p_{\r_2} p_{\r_3} p_{\r_4},}
\label{Gammahh}
\end{array}
\end{equation}
where
\newpage

    \begin{align}
    &\phantom{xxx}\nonumber\\[2cm]
       & P_1(p)=\frac{i (-p^2)^{D/2-2 } \pi^{(1-D)/2} \Gamma(1 -  \frac{D}{2} ) \Gamma(\frac{D}{2} ) }{4^{3 +  D}  \Gamma\bigl(\frac{D+3}{2} \bigr)}\Bigg\{\frac{88 + D \biggl(26 + D \Bigl( D \bigl(24 + (13 - 4 D) D\bigr)-127 \Bigr)\biggr) }{(D-2)^2}  +\nonumber\\
		&+32  \frac{ \frac{\gamma_{1}^4 (\gamma_{1} + D \gamma_{2})^2}{\alpha_{1}^2 \alpha_{2}^2} + (1 + D) \bigl(\frac{(5 - 2 D) D \gamma_{1}^2}{ D-2} + (8 - 6 D) \rho_{1}{}\bigr)}{D^2 \alpha^2 \gamma_{1}^2}-  \frac{4 (D-2) (1 + D) \bigl( D (15 + D)-28\bigr) \rho_{1}^2}{\gamma_{1}^4}+\nonumber\\
		&+ \frac{4  (1 + D) \biggl(48 + D \Bigl( D \bigl(161 + D ( 4 D-47 )\bigr)-182 \Bigr)\biggr) \rho_{1}{}}{(D-2)\gamma_{1}^2}- \frac{16 \gamma_{1} (\gamma_{1} + D \gamma_{2})}{\alpha\, \alpha_1 \,\alpha_2}\Bigg\},\nonumber\\[1.5cm]
		&P_2(p)=\frac{i \left(-p^2\right)^{D/2-2 }   \Gamma(1 -  \frac{D}{2} ) \Gamma(\frac{D}{2} )}{4^{3 +  D}  \pi^{D/2 }} \Bigg\{- \frac{2^{5 + D} \Gamma\left(\frac{D}{2} \right)}{\left(D-1\right) D \,\Gamma\left(D-2 \right)} + \nonumber\\
		&+\frac{\pi^{1/2}}{\left(\gamma_{1}^2\,D^2\left(D-2 \right)  \alpha\, \alpha_{1} \,\alpha_{2}\right)^2 \Gamma\left(\frac{D+3}{2} \right) }\Bigg[ 16 \left(D-2 \right)^2 D^3 \left(4 + 3 D\right) \alpha \,\alpha_1 \alpha_2 \gamma_1^5 \left(\gamma_{1} + D \gamma_{2}\right) +\nonumber\\
		&+ 32 \left(D\left(D-2 \right) \gamma_1^3\left(\gamma_{1} + D \gamma_{2}\right)\right)^2 -  \alpha_1^2 \alpha_2^2 \Bigg[\gamma_1^4\Big(32 \left(D-2 \right) \left(1 + D\right) \left( D \left(76 + \left( D-28 \right) D\right)-32 \right) + \nonumber\\
		&+\alpha^2\Big( D \left(640 + D \left(  D \left(2856 + D \left(1668 + D \left( D \left(1371 + D \left( D \left( 4 D-37 \right)-100 \right)\right)-3382 \right)\right)\right)-3424\right)\right)+\nonumber\\
		&+512\Big) \Big)+4\left(D-2 \right) \left(1 + D\right) \Bigg( \Bigg( D \Big(128 + D \Big(3040 + D ( D \left(3748 + D \left( D \left(79 + 3 D\right)-1006 \right)\right)-\nonumber\\
		&-5576 )\Big)\Big)-512 \Bigg) \alpha^2- 16 \left(D-2 \right)\Big( D \left(28 + \left( D-4\right) D\right)-32 \Big)\Bigg) \gamma_1^2 \rho_{1}{} +  \nonumber\\
		&+4 (\alpha \rho_{1})^2\left(D-2 \right)^3 \left(1 + D\right) \Bigg( D \Big(64 + D \Big(1168 + D \Big( D (636 +D \left(9 D-133 \right))-1404 \Big)\Big)\Big)-256\Bigg) \Bigg]\Bigg]\Bigg\},\nonumber\\
		\nonumber\\
\nonumber
	\end{align}
\newpage
$\phantom{xxx}$
\vspace{3cm}
\begin{align}
        &P_3(p)=\frac{i (-p^2)^{D/2-2 } \Gamma(1 -  \frac{D}{2} ) \Gamma(\frac{D}{2} ) }{4^{3 +  D}\pi^{D/2} }\Bigg\{\frac{2^{4 + D}  \Gamma(\frac{D}{2} )}{(D-1) \Gamma(D-2)} +\nonumber\\
		&+ \frac{\pi^{1/2} }{D^3((D-2) \alpha\,\alpha_{1}\, \alpha_{2} \gamma_{1}^2)^2 \Gamma\bigl(\frac{D+3}{2} \bigr) }\Bigg[-16 (D-2)^2 D^2 \bigl(2 + D (2 + D)\bigr) \alpha\, \alpha_1 \,\alpha_2 \gamma_{1}^5 (\gamma_{1} + D \gamma_{2}) - \nonumber\\
		&-32  D (2 + D) (\gamma_{1}^3(D-2) (\gamma_{1} + D \gamma_{2}))^2 + \alpha_{1}^2 \alpha_{2}^2 \Bigg[\Bigg[32 (D-2) (1 + D) \Bigl( D \bigl(76 + D ( 5 D-38 )\bigr)-32 \Bigr) +\nonumber\\
		&+ \Bigg[512 + D \Bigg(640 + D \Big(  D \left(3128 + D \Biggl(1296 + D \biggl(D \Bigl(1747 + D \bigl( D (4 D-5 )-302 \bigr)\Bigr)-3400 \biggr)\Biggr)\right)-\nonumber\\
		&-3552\Big)\Bigg)\Bigg]\alpha^2 \Bigg]\gamma_{1}^4  - 4 (D-2) (1 + D)\gamma_{1}^2 \rho_{1}{} \Bigg(16 (D-2) \Bigl( D \bigl(36 + ( D-10) D\bigr)-32 \Bigr) + \nonumber\\
		&+( D-4) \alpha^2\Biggl( D^2 \biggl(776 + D \Bigl( D \bigl(836 + D ( 17 D-213 )\bigr)-1312 \Bigr)\biggr)-128 \Biggr) \Bigg)  +\nonumber\\
		&+ 4 (D-2)^3 (1 + D)\alpha^2 \rho_{1}^2 \left(D \Biggl(64 + D \biggl(1168 + D \Bigl( D \bigl(666 + D ( 9 D-131 )\bigr)-1460 \Bigr)\biggr)\Biggr)-256 \right) \Bigg]\Bigg]\Bigg\},
        \nonumber\\[1.5cm]	
        &P_4(p)=- \frac{i (-p^2)^{D/2-2 } \pi^{(1-D)/2} \Gamma(1 -  \frac{D}{2} ) \Gamma(\frac{D}{2} ) }{4^{3 + D}\left(D (D-2)  \alpha\gamma_{1}^2\right)^2\,\alpha_{1}\,\alpha_{2}\Gamma\bigl(\frac{D+3}{2} \bigr)  }\Bigg\{ 16 \alpha  \gamma _1^5 (D-2)^2 \left(\gamma _1+\gamma _2 D\right) D (D(D-2) -2)-\nonumber\\
        &-32 \alpha_1 \alpha_2 \gamma _1^2 (D-2)^2 \Bigg(\frac{\gamma _1^4 D \left(\gamma _1+\gamma _2 D\right){}^2}{\alpha_1^2 \alpha_2^2}+(D+1) \Big(\frac{\gamma _1^2 D (2 D-5)}{D-2}+2 (3 D-4) \rho _1\Big)\Bigg)-\nonumber\\
        &-\alpha ^2 \alpha_1 \alpha_2 D^2 \Bigg[4 (D+1) (D (D+15)-28) (D-2)^3 \rho _1^2+{\gamma _1^4}D \Bigg(D \Big(D (D (4 D-13)-24)+127\Big)-\nonumber\\
        &-26\Bigg)-\gamma _1^24 (D+1) \Bigg(D \Big(D \Big(D (4 D-47)+161\Big)-182\Big)+48\Bigg) (D-2) \rho _1-88{\gamma _1^4}\Bigg]\Bigg\},\nonumber
\end{align}
\newpage
	\begin{align}
		&P_5(p)=- \frac{i (-p^2)^{D/2-2 } \pi^{(1-D)/2} \Gamma(1 -  \frac{D}{2} ) \Gamma(\frac{D}{2} ) }{4^{3 + D} (D-2) (D \alpha\,\alpha_{1}\,\alpha_{2}\gamma_{1}^2)^2 \Gamma\bigl(\frac{D+3}{2} \bigr) }\Bigg\{ (4D-8)^2 D \bigl( D( D-4 )-4 \bigr) \alpha \alpha_{1} \alpha_{2} \gamma_{1}^5 (\gamma_{1} + D \gamma_{2}) -\nonumber\\
		&-32 ((D-2)\gamma_{1}^3 (\gamma_{1} + D \gamma_{2}))^2 D  + \alpha_{1}^2 \alpha_{2}^2 \Bigg[\gamma_{1}^4\Bigg[32 (D-2) (1 + D) \bigl(16 + 5 ( D-4 ) D\bigr) + \nonumber\\
		&+\alpha^2\Bigg[ D \Bigg( D \left(1792 + D \Biggl( D \biggl( D \Bigl(1303 + D \bigl( D (27 + 4 D)-408 \bigr)\Bigr)-1162 \biggr)-840 \Biggr)\right)-448\Bigg) -\nonumber\\
		&-256 \Bigg] \Bigg] - 4 (D-2) (1 + D) \Bigg(16 (D-2) \bigl(8 + ( D-8) D\bigr) + \Biggl( ( D-4) D \biggl(16 - D \Bigl(360 -\nonumber\\
		&- D \bigl(434 + D ( 29 D-199 )\bigr)\Bigr)\biggr)-256 \Biggr) \alpha^2\Bigg) \gamma_{1}^2 \rho_{1}{} + 4 (D-2)^3 (1 + D) \Biggl(128 +\nonumber\\
		&+ D \biggl(32 + D \Bigl( D \bigl(444 + D ( 9 D-113)\bigr)-512 \Bigr)\biggr)\Biggr) \alpha^2 \rho_{1}^2\Bigg]\Bigg\}.\nonumber
	\end{align}
For $\Gamma^{(fh)}_{\r_1\r_2}(p)$, we have obtained the following value
\begin{equation}
\Gamma^{(fh)}_{\r_1\r_2}(p)=Q_1(p)\,(p^2)^2 \eta_{\r_1\r_2}+Q_2(p)\,p^2 p_{\r_1} p_{\r_2},
\label{Gammafh}
\end{equation}
where
\begin{equation}
\begin{array}{l}
{Q_1(p)=i\, \dfrac{2^{-(2+D)} \pi^{-D/2}(-p^2)^{-2+D/2}\Gamma\left(1-D/2\right)(\Gamma\left(D/2\right))^2}{\alpha (-1+D) D \Gamma\left(D-2\right)}}\\[4pt]
{Q_2(p)=-i\, \dfrac{2^{-(2+D)} \pi^{-D/2}(-p^2)^{-2+D/2}\Gamma\left(1-D/2\right)(\Gamma\left(D/2\right))^2}{\alpha (-1+D) \Gamma\left(D-2\right)}.}
\end{array}
\label{Qobjects}
\end{equation}
Not surprisingly, $\Gamma^{(hf)}_{\r_1\r_2}(p)=\Gamma^{(fh)}_{\r_1\r_2}(p)$.

\subsection{Two simplifying gauges}

Let us present two gauge choices for which $\Gamma^{(hh)}_{\r_1\r_2\r_3\r_4}(p)$ simplifies greatly. These gauge choices simplify  as much as possible the involved tensor structure of the {free-graviton} propagator given in (\ref{propsymbols}) and (\ref{Gpropaone}). Let us first point out that the last two summands -the most involved ones-- of that tensor structure cannot {be} set to zero at the same time. Actually, this is needed to obtain Newton's law as seen in {Appendix} C.

Let {us} set $\alpha^2=\dfrac{(2 D-4)}{D^2}$ and $\rho_1=\dfrac{1}{4} \gamma_1^2$. This choice sets to zero the second and third summands of $G^{(hh)}_{\m_1,\m_2,\m_3,\m_4}(p)$ in (\ref{Gpropaone}).
Let us further impose $\gamma_2=-\dfrac{\gamma_1}{D}$. Then, $P_1(p), P_2(p), P_3(p)$ and $P_4(p)$ in (\ref{Gammahh}) read
\begin{equation*}
\begin{array}{l}
{P_1(p)=i\,\left(256+D(-864+D(-260+D(549+5D(-40+3D))))\right)\,\Phi(p,D), }\\[8pt]
{P_2(p)=-i\,\dfrac{1}{D^4}(-4096 + D (-3072 +D (26112 +D (-19200 +D (-15664 +D (19812}\\[4pt]
{\phantom{P_2(p)=-i\,\dfrac{1}{D^4}(-4096 + D }
 + D (-7520 + D (1247 + 3 D (-44 + 3 D))))))))) \,\Phi(p,D),}\\[8pt]
 {P_3(p)=i\dfrac{1}{D^3} (-4096 + D (-3072 +D (26624+D (-21888 +D (-12912}\\[4pt]
 {\phantom{P_3(p)=i\dfrac{1}{D^3}\quad\quad}
  +D (21932 + D (-10202 + 3 D (709 + D (-70 + 3 D))))))))) \,\Phi(p,D),}\\[8pt]
{P_4(p)=-i\,(256 + D (-864 + D (-260 + D (549 + 5 D (-40 + 3 D)))))\,\Phi(p,D),}\\[8pt]
{P_5(p)=-i\dfrac{1}{D^2}(2048 + D (2560 +D (-11776 + 3 D (1472}\\[4pt]
{\phantom{P_2(p)=-i\,\dfrac{1}{D^4}}
+ D (2544 + D (-2284 + D (689 + D (-80 + 3 D))))))))\,\left(D-2\right)\Phi(p,D),}
\end{array}
\end{equation*}
where
\begin{equation*}
\Phi(p,D)=\dfrac{4^{-(4+D)}\pi^{(1- D)/2}(-p^2)^{-2 + D/2}\Gamma\left(1-D/2\right)\Gamma\left(D/2\right)}{\left(D-2\right)\Gamma\left(3+D/2\right)}.
\end{equation*}

The second gauge choice sets to zero the last summand of {$G^{(hh)}_{\m_1 \m_2 \m_3 \m_4}(p)$} in (\ref{Gpropaone}) and reads
\begin{equation*}
 \alpha^2=\dfrac{(2 D-4)}{D^2},\, \rho_1=-\dfrac{1}{2(D-2)} \gamma_1^2\quad\text{and}\quad \gamma_2=-\dfrac{\gamma_1}{D}.
\end{equation*}
For this gauge choice,  $P_1(p), P_2(p), P_3(p)$ and $P_4(p)$ in (\ref{Gammahh}) are given by the following expressions
\begin{equation*}
\begin{array}{l}
{P_1(p)=-i\,(-32 + D (74 + D (51 + D (-37 + 6 D))))\Psi(p,D), }\\[8pt]
{P_2(p)=-i\,\dfrac{1}{D^4}(-512 + D (-640 +D (3040}\\[4pt]
{ \phantom{P_2(p)=-i\,\dfrac{1}{1}}
+D (-984 + D (-2468 + D (1498 + D (-213 + D (-13 + 2 D))))))))\,\Psi(p,D),}\\[8pt]
 {P_3(p)=i\,\dfrac{1}{D^3}(-512 + D (-640 + D (3104 + D (-1304}\\[4pt]
 {\phantom{P_3(p)=i\,\dfrac{1}{D^3}(-512}
  + D (-2256 + D (1816 + D (-431 + D (23 + 2 D))))))))\,\Psi(p,D),}\\[8pt]
{P_4(p)=i\, (-32 + D (74 + D (51 + D (-37 + 6 D))))\,\Psi(p,D),}\\[8pt]
{P_5(p)=-i\,\dfrac{1}{D^2}(256 + D (448 +D (-1296}\\[4pt]
{\phantom{ P_5(p)=-i\,\dfrac{1}{D^2}}
+ D (-16 + D (986 + D (-449 + D (51 + 2 D)))))))\,\left(D-2\right)\Psi(p,D),}
\end{array}
\end{equation*}
where
\begin{equation*}
\Psi(p,D)=\dfrac{2^{-(5+2D)}\pi^{(1- D)/2}(-p^2)^{-2 + D/2}\Gamma\left(1-D/2\right)\Gamma\left(D/2\right)}{\left(D-2\right)\Gamma\left(3+D/2\right)}.
\end{equation*}

\section{Comments and conclusions.}

We have {spelt} out a BRST formalism where quantum corrections, in any number of dimensions and around a Minkowski background, can be computed for unimodular gravity with Weyl invariance. We have applied the formalism to the one-loop computation of the complete graviton propagator for a large family of gauge-fixing terms and in any number of dimensions. Our main conclusion is that computing quantum gravitational corrections in  unimodular gravity is feasible. Of course, the computations are very involved due to the fact that the free graviton propagator contains {higher-order} poles.

{
\section{Acknowledgements.}
    We are thankfull to Professor Enrique \'Alvarez V\'azquez for plenty of illuminating blackboard discussions.
    One of us (EVA) is grateful for informative email exchange with M. Herrero-Valea and acknowledges partial financial support by the Spanish MINECO through the Centro de excelencia Severo Ochoa Program under Grant CEX2020-001007-S funded by MCIN/AEI/10.13039/501100011033 as well as to the European Union's Horizon 2020 research and innovation programme under the Marie Sklodowska-Curie grant agreement No 860881-HIDDeN and also byGrant PID2019-108892RB-I00 funded by MCIN/AEI/ 10.13039/501100011033 and by ``ERDF A way of making Europe''.
}

\appendix

\section{ Results when D is close to 4.}

Here we shall display the values of $\Gamma^{(hh)}_{\r_1\r_2\r_3\r_4}(p)$ and  $\Gamma^{(fh)}_{\r_1\r_2}(p)$ in the vicinity of $D=4$. We shall choose the gauges in the subsection 4.1. Let us  begin with the gauge choice  $\alpha^2=\dfrac{(2 D-4)}{D^2},\, \rho_1=\dfrac{1}{4} \gamma_1^2$ and $\gamma_2=-\dfrac{\gamma_1}{D}$. For this gauge choice, $P_1(p),\, P_2(p),\, P_3(p)$ and $P_4(p)$ in (\ref{Gammahh}) read
\begin{equation*}
\begin{array}{l}
{P_1(p)=-i\,\dfrac{21}{320\pi^2}\big\{\dfrac{1}{D-4}+\dfrac{1}{2}{\rm Ln}\left(-\frac{p^2}{4\pi e^{-\gamma}\mu^2}\right)\big\}+i\,\dfrac{813}{12800\pi^2}+O(D-4),}\\[8pt]
{P_2(p)=i\,\dfrac{37}{960\pi^2}\big\{\dfrac{1}{D-4}+\dfrac{1}{2}{\rm Ln}\left(-\frac{p^2}{4\pi e^{-\gamma}\mu^2}\right)\big\}-i\,\dfrac{13691}{230400\pi^2}+O(D-4),}\\[8pt]
 {P_3(p)=-i\,\dfrac{11}{480\pi^2}\big\{\dfrac{1}{D-4}+\dfrac{1}{2}{\rm Ln}\left(-\frac{p^2}{4\pi e^{-\gamma}\mu^2}\right)\big\}+i\,\dfrac{3779}{57600\pi^2}+O(D-4),}\\[8pt]
{P_4(p)=i\,\dfrac{21}{320\pi^2}\big\{\dfrac{1}{D-4}+\dfrac{1}{2}{\rm Ln}\left(-\frac{p^2}{4\pi e^{-\gamma}\mu^2}\right)\big\}-i\,\dfrac{813}{12800\pi^2}+O(D-4),}\\[8pt]
{P_5(p)=-i\,\dfrac{41}{240\pi^2}\big\{\dfrac{1}{D-4}+\dfrac{1}{2}{\rm Ln}\left(-\frac{p^2}{4\pi e^{-\gamma}\mu^2}\right)\big\}+i\,\dfrac{1169}{28800\pi^2}+O(D-4).}
\end{array}
\end{equation*}

When the gauge parameters satisfy $\alpha^2=\dfrac{(2 D-4)}{D^2},\, \rho_1=-\dfrac{1}{2(D-2)} \gamma_1^2\quad\text{and}\quad \gamma_2=-\dfrac{\gamma_1}{D}$, we have
\begin{equation*}
\begin{array}{l}
{P_1(p)=-i\,\dfrac{31}{1920\pi^2}\big\{\dfrac{1}{D-4}+\dfrac{1}{2}{\rm Ln}\left(-\frac{p^2}{4\pi e^{-\gamma}\mu^2}\right)\big\}+i\,\dfrac{1967}{115200\pi^2}+O(D-4),}\\[8pt]
{P_2(p)=i\,\dfrac{17}{960\pi^2}\big\{\dfrac{1}{D-4}+\dfrac{1}{2}{\rm Ln}\left(-\frac{p^2}{4\pi e^{-\gamma}\mu^2}\right)\big\}-i\,\dfrac{1583}{115200\pi^2}+O(D-4),}\\[8pt]
 {P_3(p)=-i\,\dfrac{1}{80\pi^2}\big\{\dfrac{1}{D-4}+\dfrac{1}{2}{\rm Ln}\left(-\frac{p^2}{4\pi e^{-\gamma}\mu^2}\right)\big\}+i\,\dfrac{53}{3200\pi^2}+O(D-4),}\\[8pt]
{P_4(p)=i\,\dfrac{31}{1920\pi^2}\big\{\dfrac{1}{D-4}+\dfrac{1}{2}{\rm Ln}\left(-\frac{p^2}{4\pi e^{-\gamma}\mu^2}\right)\big\}-i\,\dfrac{1967}{115200\pi^2}+O(D-4),}\\[8pt]
{P_5(p)=-i\,\dfrac{19}{160\pi^2}\big\{\dfrac{1}{D-4}+\dfrac{1}{2}{\rm Ln}\left(-\frac{p^2}{4\pi e^{-\gamma}\mu^2}\right)\big\}-i\,\dfrac{209}{3200\pi^2}+O(D-4).}
\end{array}
\end{equation*}

$Q_1(p)$ and $Q_2(p)$ in (\ref{Qobjects}) have the following Laurent expansion at $D=4$:
\begin{equation*}
\begin{array}{l}
{Q_1(p)=i\,\dfrac{1}{384\alpha\pi^2}\big\{\dfrac{1}{D-4}+\dfrac{1}{2}{\rm Ln}\left(-\frac{p^2}{4\pi e^{-\gamma}\mu^2}\right)\big\}-i\,\dfrac{13}{4608\alpha\pi^2}+O(D-4),}\\[8pt]
{Q_2(p)=-i\,\dfrac{1}{96\alpha\pi^2}\big\{\dfrac{1}{D-4}+\dfrac{1}{2}{\rm Ln}\left(-\frac{p^2}{4\pi e^{-\gamma}\mu^2}\right)\big\}+i\,\dfrac{5}{576\alpha\pi^2}+O(D-4).}
 \end{array}
\end{equation*}

In the previous result $\gamma$ stands for the Euler's constant and $\mu$ for the dimensional regularization scale.

Let us finally point out that in the minimal {subtraction} scheme of dimensional regularization the renormalized one-loop correction to the graviton propagator is obtained by removing first the single pole at $D=4$ from the $P_i(p)$'s and $Q_i(p)$'s above, then, inserting the resulting renormalized $P_i(p)$'s and $Q_i(p)$'s in the corresponding formulae in  (\ref{Gammahh}) and (\ref{Gammafh}) and, finally, substituting the so renormalized $\Gamma^{(hh)}_{\r_1\r_2\r_3\r_4}(p)$ and  $\Gamma^{(fh)}_{\r_1\r_2}(p)$ in (\ref{oneloopprop}), as $D$ goes to 4.

\section{Value of each Feynman diagram.}

Each diagram in Figure 3 is given by the following general tensor structure
\begin{equation*}
\begin{array}{l}
{R_1(p)\,(p^2)^2\left(\eta_{\r_1\r_3}\eta_{\r_2\r_4}+\eta_{\r_1\r_4}\eta_{\r_2\r_3}\right)
+R_2(p)\,(p^2)^2\eta_{\r_1\r_2}\eta_{\r_3\r_4}
+R_3(p)\,p^2\left(\eta_{\r_3\r_4}p_{\r_1} p_{\r_2}+\eta_{\r_1\r_2}p_{\r_3} p_{\r_4}\right)+}\\[4pt]
{R_4(p)\,p^2\left(\eta_{\r_1\r_3}p_{\r_2} p_{\r_4}+\eta_{\r_1\r_4}p_{\r_2} p_{\r_3}+\eta_{\r_2\r_3}p_{\r_1} p_{\r_4}+\eta_{\r_2\r_4}p_{\r_1} p_{\r_3}\right)+R_5(p)\, p_{\r_1} p_{\r_2} p_{\r_3} p_{\r_4},}
\end{array}
\end{equation*}
where the value of each $R_i(p)$, $i=1...5$, depends on the diagram. Below we give the value of $R_i(p)$, $i=1..5$, for each diagram in such figure.

\noindent
{\bf\underline{ Diagram 1 in figure 3.}}
\vspace{1cm}
\begin{equation*}
\begin{array}{l}
{R_1(p)=\frac{i \pi^{\frac{1-D}{2}}(-p^2)^{-2+\frac{D}{2}}\Gamma(1-\tfrac{D}{2} ) \Gamma(\tfrac{D}{2} )}{4^{3 + D} (-2 + D)^2 D^2 \alpha^2 \gamma_1^4 \Gamma\bigl(\tfrac{D+3}{2}\bigr)}f_1[D,\a,\rho_1\g_1],}
\end{array}
\end{equation*}
with
\begin{align}
&f_1[D,\a,\rho_1\g_1]=D \Bigg\{-320-32D\bigl(1 + D(-7+2 D)\bigr) +\nonumber\\
&+ D \Bigg[120 + D \biggl(34 + D \Bigl(-167 + D \bigl(42 + (11 - 4 D) D\bigr)\Bigr)\biggr)\Bigg] \alpha^2\Bigg\}\gamma_1^4 +\nonumber\\
&+ 4 (-2 + D) (1 + D) \Bigg\{-128 +160D -48D^2 +\nonumber\\
&+ D^2\biggl(48 +D \Bigl(-182 + D \bigl(161 + D (-47 + 4 D)\bigr)\Bigr)\biggr) \alpha^2\Bigg\} \gamma_1^2 \rho_1 -\nonumber\\
&- 4 (-2 + D)^3 D^2 (1 + D) \bigl(-28 + D (15 + D)\bigr) \alpha^2 \rho_1^2;
\nonumber
\end{align}
\newpage
\begin{equation*}
\begin{array}{l}
{R_2(p)=-\frac{i \pi^{\frac{1-D}{2}}(-p^2)^{-2+\frac{D}{2}}\Gamma(1-\tfrac{D}{2} ) \Gamma(\tfrac{D}{2} )}{4^{3 + D} (-2 + D)^2 D^4 \alpha^2 \gamma_1^4 \Gamma\bigl(\tfrac{D+3}{2}\bigr)}f_2[D,\a,\rho_1\g_1],}
\end{array}
\end{equation*}
with
\begin{align}
&f_2[D,\a,\rho_1\g_1]=\Bigg\{32 (-2 + D) (1 + D) \Bigl(-32 + D \bigl(76 + (-28 + D) D\bigr)\Bigr) + \nonumber\\
&+\Bigg[512 + 640D + D^2 \Big[-3424 + D \Big(2856 + D \big(1700 +\nonumber\\
&+ D \left[-3326 + D \left(1363 + D (-150 + D (-19 + 4 D))\right)\right]\big)\Big)\Big]\Bigg]\alpha^2\Bigg\} \gamma_1^4 + \nonumber\\
&+4 (-2 + D) (1 + D) \Bigg\{-16 (-2 + D) \Big[-32 + D \bigl(28 + (-4 + D) D\bigr)\Big] +\nonumber\\
&+ \Bigg[-512 + 128D + 3040D^2 + \nonumber\\
&+D^3 \left[-5576 + D \Bigl(3748 + D \bigl(-1006 + D (79 + 3 D)\bigr)\Bigr)\right]\Bigg]\alpha^2\Bigg\}\gamma_1^2 \rho_1 +\nonumber\\
&+ 4 (-2 + D)^3 (1 + D) \Bigg\{-256 + 64D + \nonumber\\
&+D^2 \Big(1168 + D \left[-1404 + D \left(636 + D (-133 + 9 D)\right)\right]\Big)\Bigg\}\alpha^2 \rho_1^2;\nonumber\\
\nonumber
\end{align}
\newpage
\begin{align}
R_3(p)=-\frac{i \pi^{\frac{1-D}{2}}(-p^2)^{-2+\frac{D}{2}}\Gamma(1-\tfrac{D}{2} ) \Gamma(\tfrac{D}{2} )}{4^{3 + D} (-2 + D)^2 D^3 \alpha^2 \gamma_1^4 \Gamma\bigl(\tfrac{D+3}{2}\bigr)}f_3[D,\a,\rho_1\g_1],
\nonumber
\end{align}
with
\begin{align}
&f_3[D,\a,\rho_1\g_1]=\Bigg\{-32 (-2 + D) (1 + D) \Bigl(-32 + D \bigl(76 + D (-38 + 5 D)\bigr)\Bigr) - \nonumber\\
&- \Bigg[512 + 640D + D^2 \Bigg(-3552 + D \Big[3128 + D \Big(1344+ \nonumber \\
& + D \biggl(-3408 + D \left[1695 + D \bigl(-268 + D (-11 + 4 D)\bigr)\right]\biggr)\Big)\Big]\Bigg)\Bigg] \alpha^2\Bigg\}\gamma 1^4+  \nonumber \\
&+ 4 (-2 + D) (1 + D) \Bigg\{16 (-2 + D) \Bigl(-32 + D \bigl(36 + (-10 + D) D\bigr)\Bigr)+ \nonumber\\
& +(4-D) \Bigg[128 - D^2 \Biggl(776 + D \big[-1312 + D \bigl(836+ D ( 17 D-213)\bigr)\big]\Biggr)\Bigg] \alpha^2\Bigg\}\gamma 1^2 \rho_1- \nonumber \\
&  - 4 (-2 + D)^3 (1 + D) \Bigg\{-256 + 64D +\nonumber\\
&+ D^2\biggl(1168  + D \Bigl(-1460 + D \bigl(666 + D (-131 + 9 D)\bigr)\Bigr)\biggr)\Bigg\}\alpha^2 \rho_1^2 ;\nonumber\\
\nonumber
\end{align}
\newpage
\begin{align}
R_4(p)=-\frac{i \pi^{\frac{1-D}{2}}(-p^2)^{-2+\frac{D}{2}}\Gamma(1-\tfrac{D}{2} ) \Gamma(\tfrac{D}{2} )}{4^{3 + D} (-2 + D)^2 D^2 \alpha^2 \gamma_1^4 \Gamma\bigl(\tfrac{D+3}{2}\bigr)}f_4[D,\a,\rho_1\g_1],
\nonumber
\end{align}
with
\begin{align}
&f_4[D,\a,\rho_1\g_1]=- D \Bigg\{32 (-2 + D) (1 + D) (-5 + 2 D) +\nonumber\\
&+ D \Biggl[-152 + D \biggl(6 + D \Bigl(183 + D \bigl(-80 + D (5 + 2 D)\bigr)\Bigr)\biggr)\Biggr] \alpha^2\Bigg\}\gamma 1^4 +\nonumber\\
&+ 4 (-2 + D) (1 + D) \Bigg\{-128 + 160D - 48 D^2 +\nonumber\\
&+ D^2 \Biggl(48 + D \bigl(-182 + D \bigl(161 + D (-47 + 4 D)\bigr)\bigr)\Biggr) \alpha^2\Bigg\}\gamma 1^2 \rho_1 -\nonumber\\
&- 4 (-2 + D)^3 D^2 (1 + D) \bigl(-28 + D (15 + D)\bigr) \alpha^2 \rho_1^2 ;
\nonumber
\end{align}

\begin{align}
R_5(p)=-\frac{i \pi^{\frac{1-D}{2}}(-p^2)^{-2+\frac{D}{2}}\Gamma(1-\tfrac{D}{2} ) \Gamma(\tfrac{D}{2} )}{4^{3 + D} (-2 + D)^2 D^2 \alpha^2 \gamma_1^4 \Gamma\bigl(\tfrac{D+3}{2}\bigr)}f_5[D,\a,\rho_1\g_1],
\nonumber
\end{align}
with
\begin{align}
&f_5[D,\a,\rho_1\g_1]=\Bigg\{32 (-2 + D) (1 + D) \bigl(16 + 5 (-4 + D) D\bigr) +\nonumber\\
&+ \Bigg[-256 + D \Big(-448 + D\big[1920 + D \Big(-872 + D \big(-1266 + D \big(1375 +\nonumber \\
& + D \bigl(-426 + D (29 + 4 D)\bigr)\big)\big)\Big)\big]\Big)\Bigg] \alpha^2\Bigg\} \gamma_1^4 -\nonumber\\
&- 4 (-2 + D) (1 + D) \Bigg\{-256 (1 + \alpha^2) + (-4 + D) D \Biggl(16 (-6 + D) + \nonumber \\
&+ \biggl(16 + D \Bigl(-360 + D \bigl(434 + D (-199 + 29 D)\bigr)\Bigr)\biggr) \alpha^2\Biggr)\Bigg\} \gamma 1^2 \rho_1+ \nonumber \\
& + 4 (D-2)^3 (1 + D) \Bigg\{128 + 32D + D^2 \Bigl(-512 + D \bigl(444 + D (9 D-113)\bigr)\Bigr)\Bigg\} \alpha^2 \rho_1^2.\nonumber\\
\nonumber
\end{align}
\noindent
{\bf\underline{ Diagram 2 in figure 3.}}
\begin{align}
&R_1(p)=\frac{i \bigl(-4 + (-5 + D) D\bigr) \pi^{\frac{1-D}{2}}(-p^2)^{-2+\frac{D}{2}}\Gamma(1-\tfrac{D}{2} ) \Gamma(\tfrac{D}{2} )}{2^{5 + 2D}  \Gamma\bigl(\tfrac{3 + D}{2} \bigr)},\nonumber\\
&R_2(p)=  \frac{i \bigl(4 + D (11 + 9 D)\bigr) \pi^{\frac{1-D}{2}}(-p^2)^{-2+\frac{D}{2}} \Gamma(1-\tfrac{D}{2} ) \Gamma(\tfrac{D}{2} )}{2^{5 + 2D}   \Gamma\bigl(\tfrac{3 + D}{2} \bigr)},\nonumber\\
&R_3(p)= \frac{i D \bigl(-6 + (-5 + 3 D) D\bigr) \pi^{\frac{1-D}{2}}(-p^2)^{-2+\frac{D}{2}}\Gamma(1-\tfrac{D}{2} ) \Gamma(\tfrac{D}{2} )}{2^{5 + 2D} \Gamma\bigl(\tfrac{3 + D}{2} \bigr)},\nonumber\\
&R_4(p)= \frac{i \bigl(8 + (-4 + D) (-1 + D) D\bigr)\pi^{\frac{1-D}{2}} (-p^2)^{-2+\frac{D}{2}}\Gamma(1-\tfrac{D}{2} ) \Gamma(\tfrac{D}{2} )}{2^{5+ 2D} \Gamma\bigl(\tfrac{3 + D}{2} \bigr)},\nonumber\\
&R_5(p)=\frac{i (-2 + D) \Bigl(16 + D \bigl(12 + (-5 + D) D\bigr)\Bigr) \pi^{\frac{1-D}{2}} (-p^2)^{-2+\frac{D}{2}}\Gamma(1-\tfrac{D}{2} ) \Gamma(\tfrac{D}{2} )}{2^{5 +2 D}  \Gamma\bigl(\tfrac{3 + D}{2} \bigr)}.
\nonumber
\end{align}
\noindent
{\bf\underline{ Diagram 3 in figure 3.}}
\begin{align}
&R_1(p)=\frac{i \gamma_1^2 (\gamma_1 + D \gamma_2)^2 \pi^{\frac{1-D}{2}}(-p^2)^{-2+\frac{D}{2}}\Gamma(1-\tfrac{D}{2} ) \Gamma(\tfrac{D}{2} )}{2^{1+ 2D} D^2  \alpha^2 \alpha_1^2 \alpha_2^2 \Gamma\bigl(\tfrac{D+3}{2} \bigr)},\nonumber\\
&R_2(p)=\frac{i\gamma_1^2 (\gamma_1 + D \gamma_2)^2 \pi^{\frac{1-D}{2}}(-p^2)^{-2+\frac{D}{2}}\Gamma(1-\tfrac{D}{2} ) \Gamma(\tfrac{D}{2} )}{2^{1 +2 D} D^2   \alpha^2 \alpha_1^2 \alpha_2^2 \Gamma\bigl(\tfrac{D+3}{2} \bigr)},\nonumber\\
&R_3(p)=- \frac{i (2 + D)  \gamma_1^2 (\gamma_1 + D \gamma_2)^2  \pi^{\frac{1-D}{2}}(-p^2)^{-2+\frac{D}{2}}\Gamma(1-\tfrac{D}{2} ) \Gamma(\tfrac{D}{2} )}{2^{1+ 2D} D^2  \alpha^2 \alpha_1^2 \alpha_2^2 \Gamma\bigl(\tfrac{D+3}{2} \bigr)},\nonumber\\
&R_4(p)=\frac{i \gamma_1^2 (\gamma_1 + D \gamma_2)^2  \pi^{\frac{1-D}{2}}(-p^2)^{-2+\frac{D}{2}}\Gamma(1-\tfrac{D}{2} ) \Gamma(\tfrac{D}{2} )}{2^{1+2 D} D \alpha^2 \alpha_1^2 \alpha_2^2 \Gamma\bigl(\tfrac{D+3}{2} \bigr)},\nonumber\\
&R_5(p)=\frac{i \gamma_1^2 (\gamma_1 + D \gamma_2)^2 \left(D-2\right)\pi^{\frac{1-D}{2}}(-p^2)^{-2+\frac{D}{2}}\Gamma(1-\tfrac{D}{2} ) \Gamma(\tfrac{D}{2} )}{2^{1 + 2D} D \alpha^2 \alpha_1^2 \alpha_2^2 \Gamma\bigl(\tfrac{D+3}{2} \bigr)}.
\nonumber
\end{align}
\newpage
{\bf\underline{ {Diagrams} 4 and 5 in figure 3.}}

Diagrams 4 and 5 have the same $R_i(p)$, $i=1..5$, which are given  next:
\begin{align}
&R_1(p)=-i\frac{ \gamma_1 (\gamma_1 + D \gamma_2) \pi^{\frac{1-D}{2}}(-p^2)^{-2+\frac{D}{2}}\Gamma(1-\tfrac{D}{2} ) \Gamma(\tfrac{D}{2} )}{2^{3+ 2D}  \alpha \alpha_1 \alpha_2\Gamma\bigl(\tfrac{D+3}{2} \bigr)},\nonumber\\
&R_(p)=i\frac{(4 + 3 D)  \gamma_1 (\gamma_1 + D \gamma_2) \pi^{\frac{1-D}{2}}(-p^2)^{-2+\frac{D}{2}}\Gamma(1-\tfrac{D}{2} ) \Gamma(\tfrac{D}{2} )}{2^{3 + 2D} D\alpha \alpha_1 \alpha_2 \Gamma\bigl(\tfrac{D+3}{2} \bigr)},\nonumber\\
&R_3(p)=-i\frac{\bigl(2 + D (2 + D)\bigr)  \gamma_1 (\gamma_1 + D \gamma_2) \pi^{\frac{1-D}{2}}(-p^2)^{-2+\frac{D}{2}}\Gamma(1-\tfrac{D}{2} ) \Gamma(\tfrac{D}{2} )}{2^{3+ 2D} D  \alpha \alpha_1 \alpha_2 \Gamma\bigl(\tfrac{D+3}{2} \bigr)},\nonumber\\
&R_4(p)=-i\frac{\bigl(-2 + (-2 + D) D\bigr)  \gamma_1 (\gamma_1 + D \gamma_2)\pi^{\frac{1-D}{2}} (-p^2)^{-2+\frac{D}{2}}\Gamma(1-\tfrac{D}{2} ) \Gamma(\tfrac{D}{2} )}{2^{3 + 2D} D \alpha \alpha_1 \alpha_2 \Gamma\bigl(\tfrac{D+3}{2} \bigr)},\nonumber\\
&R_5(p)=-i\frac{\bigl(-4 + (-4 + D) D\bigr) \gamma_1 (\gamma_1 + D \gamma_2)\left(D-2\right)\pi^{\frac{1-D}{2}}(-p^2)^{-2+\frac{D}{2}}\Gamma(1-\tfrac{D}{2} ) \Gamma(\tfrac{D}{2} )}{2^{3 + 2D} D \alpha \alpha_1 \alpha_2 \Gamma\bigl(\tfrac{D+3}{2} \bigr)}.\nonumber\\
\nonumber
\end{align}

{\bf\underline{ Diagrams 6 and 7 in figure 3.}}

Diagrams 6 and 7 have the same $R_i(p)$, $i=1..5$, whose values read:
\begin{align}
&R_1(p)=0,\nonumber\\
&R_2(p)=-i\frac{ \pi^{-\frac{D}{2}}(-p^2)^{-2+\frac{D}{2}}\Gamma(1-\tfrac{D}{2} ) \left(\Gamma(\tfrac{D}{2} )\right)^2}{ 2^{2+D} D(-1+D) \Gamma(-2 + D)},\nonumber\\
&R_3(p)=i\frac{\pi^{-\frac{D}{2}}(-p^2)^{-2+\frac{D}{2}}\Gamma(1-\tfrac{D}{2} ) \left(\Gamma(\tfrac{D}{2} )\right)^2}{2^{3+D} (-1 + D) \Gamma(-2 + D)},\nonumber\\
&R_4(p)=0,\nonumber\\
&R_5(p)=0.
\nonumber
\end{align}

\section{ Newton's Law and gauge-fixing.}

In our formulation of unimodular gravity the minimal action,$S_\phi$, of a massive scalar field, $\phi$, coupled to gravity in 4 dimensions reads
\begin{equation*}
\begin{array}{l}
{S_{\phi}=\dfrac{1}{2}\int d^4x\,\hat{g}^{\mu\nu}\partial_\mu\phi(x)\partial_\nu\phi(x)-m^2(\phi(x))^2=}\\[8pt]
{\phantom{S_{\phi}}
=\dfrac{1}{2}\int d^4x\,\eta^{\mu\nu}\partial_\mu\phi(x)\partial_\nu\phi(x)-m^2(\phi(x))^2-\dfrac{1}{2}\int d^4x\,\hat{h}_{\mu\nu}(x) T^{\mu\nu}(x)+ O((h_{\mu\nu})^2)}
\end{array}
\end{equation*}
where $\hat{g}_{\mu\nu}=\dfrac{g_{\mu\nu}}{|g|^{1/4}}$ --with $g_{\mu\nu}=\eta_{\mu\nu}+h_{\mu\nu}$,  $T^{\mu\nu}(x)$ is the Energy-Momentum tensor
\begin{equation*}
T^{\mu\nu}(x)=\partial^{\mu}\phi\partial^{\nu}\phi-\dfrac{1}{2}\eta^{\mu\nu}(\partial_\lambda\phi\partial^\lambda\phi-m^2\phi^2),
\end{equation*}
and
\begin{equation*}
\hat{h}_{\m\n}=h_{\m\n}-\dfrac{1}{4}h\,\eta_{\m\n},\quad h=h^\rho_\rho.
\end{equation*}

Following reference \cite{Donoghue:1994dn}, we shall obtain the Newtonian potential by considering two very massive scalar particles, with masses $m_1^2$ and $m_2^2$, and computing the static limit of the amplitude of the tree-level on-shell process in which those particles exchange one virtual  graviton. In our case, the amplitude, ${\cal A}$, in question is given by
\begin{equation*}
{\cal A}=-\dfrac{i}{4}T^1_{\m_1\m_2}(p_1,p'_1)\,\langle\hat{h}^{\m_1\m_2}(k)\hat{h}^{\m_3\m_4}(-k)\rangle_0\, T^2_{\m_3\m_4}(p_2,p'_2)
\end{equation*}
where $k=p_1-p'_1=p'_2-p_2$, $p_i^2=(p'_i)^2=m_i^2$, $i=1,2$ and $T^i_{\m_1\m_2}(p_i,p'_i)$ is the lowest order contributions to the on-shell energy-momentum tensor, {i.e.},
\begin{equation*}
T^i_{\m_1\m_2}(p_i,p'_i)=p_{i\,\m_1}p'_{i\,\m_2}+p_{i\,\m_2}p'_{i\,\m_1}+\dfrac{1}{2}\,k^2\eta_{\m_1\m_2}.
\end{equation*}
The Green function $\langle\hat{h}^{\m_1\m_2}(k)\hat{h}^{\m_3\m_4}(-k)\rangle_0$ is obtained in an obvious way from the graviton free propagator
$\langle h_{\m_1\m_2}(k) h_{\m_3\m_4}(-k)\rangle_0$ in (\ref{propsymbols}) and (\ref{Gpropaone}).

For very massive particles in the static limit, we have $k^\mu=(0,\vec{k})^\mu$ and $T^i_{\m_1\m_2}(p_i,p'_i)=2 m_i^2\eta_{\m_1 0}\eta_{\m_2 0}$, $i=1,2$, so that the Newtonian potential, $V_{\rm Nw}(\vec{k})$, in momentum space is given by
\begin{equation}
V_{\rm Nw}(\vec{k})=\dfrac{1}{2m_1\,2 m_2}{\cal A}=-i\dfrac{1}{4}m_1 m_2 \langle\hat{h}^{00}(k)\hat{h}^{00}(-k)\rangle_0.
\label{NWAMP}
\end{equation}
Of course, a correct quantum theory of gravity must yield
 \begin{equation}
 V_{\rm Nw}(\vec{k})\,=\,-\dfrac{1}{8}\dfrac{m_1 m_2}{\vec{k}^2},
 \label{NWPOT}
\end{equation}
which is the Fourier transform of the Newtonian potential found in any elementary textbook in the unit system $8\pi  G =1$.

Let us see what restriction agreement between (\ref{NWAMP}) and (\ref{NWPOT}) imposes on the free graviton propagator $\langle h_{\m_1\m_2}(k) h_{\m_3\m_4}(-k)\rangle_0$.
Let us introduce the following general ansatz for the free graviton propagator
\begin{equation}
\begin{array}{l}
{\langle h_{\m_1\m_2}(k) h_{\m_3\m_4}(-k)\rangle_0=}\\[4pt]
{i\,\dfrac{A_1}{k^2}\left(\eta_{\mu_1\mu_3}\eta_{\mu_2\mu_4}+\eta_{\mu_1\mu_4}\eta_{\mu_2\mu_3}-\eta_{\m_1\m_2}\eta_{\m_3\m_4}\right)
+i\,\dfrac{A_2}{k^2}\eta_{\mu_1\mu_2}\eta_{\mu_3\mu_4}
+i\,\dfrac{A_3}{(k^2)^2}\left(\eta_{\mu_3\mu_4}k_{\mu_1} k_{\mu_2}+\eta_{\mu_1\mu_2}k_{\mu_3} k_{\mu_4}\right)}\\[4pt]
{+i\, \dfrac{A_4}{(k^2)^2}\left(\eta_{\mu_1\mu_3}k_{\mu_2} k_{\mu_4}+\eta_{\mu_1\mu_4}k_{\mu_2} k_{\mu_3}+\eta_{\mu_2\mu_3}k_{\mu_1} k_{\mu_4}+\eta_{\mu_2\mu_4}k_{\mu_1} k_{\mu_3}\right)
+ i\,\dfrac{A_5}{(k^2)^3} k_{\mu_1} k_{\mu_2} k_{\mu_3} k_{\mu_4}.}
\end{array}
\label{genprop}
\end{equation}
$A_i$, $i=2..5$ are  constants. We shall assume that
\begin{equation*}
A_1=\dfrac{1}{2}
\end{equation*}
so that the contribution
\begin{equation*}
i\,\dfrac{A_1}{p^2}\left(\eta_{\mu_1\mu_3}\eta_{\mu_2\mu_4}+\eta_{\mu_1\mu_4}\eta_{\mu_2\mu_3}-\eta_{\m_1\m_2}\eta_{\m_3\m_4}\right)
\end{equation*}
to the propagator can be {interpreted} as coming from the bit of the graviton field which contains the physical graviton polarizations and the corresponding creation and annihilation operators.

Using (\ref{genprop}), one readily deduces that
\begin{equation*}
\langle\hat{h}^{00}(k)\hat{h}^{00}(-k)\rangle_0=-\dfrac{i}{4\vec{k}^2}\big[3+\dfrac{1}{4} A_5+ A_4\big],
\end{equation*}
if $k^\m=(0,\vec{k})^\mu$. Inserting the previous result in (\ref{NWAMP}) yields
\begin{equation*}
V_{\rm Nw}(\vec{k})=-\dfrac{m_1 m_2}{16\,\vec{k}^2}\big[3+\dfrac{1}{4} A_5+ A_4\big].
\end{equation*}
Hence,
\begin{equation}
A_4+\dfrac{1}{4} A_5=-1,
\label{linearconstraint}
\end{equation}
for the Newtonian potential must be given by (\ref{NWPOT}).

We thus conclude that both $A_4$ and $A_5$ in the free graviton propagator in (\ref{genprop}) cannot vanish at the same time, { regardless} of the Lorentz covariant gauge-fixing term that one uses. Notice that for our general class of gauge-fixing terms introduced in section 2, we have --see (\ref{Gpropaone})--
\begin{equation*}
A_4=-\dfrac{\gamma_1^2-4\rho_1}{2\gamma_1^2},\quad A_5=-2\dfrac{\gamma_1^2+4\rho_1}{\gamma_1^2},
\end{equation*}
which satisfy (\ref{linearconstraint}).

\end{document}